\documentclass[balance,upint,subscriptcorrection,varvw,mathalfa=cal=boondoxo,spanish,french,vietnamese,russian,greek,pdf-a,colorlinks]{asmeconf}

\usepackage{comment}

\newcommand{\p}{\partial}
\newcommand{\<}{\langle}
\renewcommand{\>}{\rangle}

\hypersetup{%
	pdfauthor={Yuanwei Bin},
	pdftitle={Large-eddy simulation of separated flows on unconventionally coarse grids},                  
	pdfsubject = {Describes the asmeconf LaTeX template},			  
}

\begin{document}




\title{Large-eddy simulation of separated flows on unconventionally coarse grids} 
 
%
%
%

\SetAuthors{%
	Yuanwei Bin\affil{1}\affil{2}, 
	George I. Park\affil{3}\CorrespondingAuthor{gipark@seas.upenn.edu}, 
	Yu Lv\affil{4}, 
	Xiang I.A. Yang\affil{1}
	}

\SetAffiliation{1}{Department of Mechanical Engineering, The Pennsylvania State University, University Park, Pennsylvania, 16802}
\SetAffiliation{2}{State Key Laboratory for Turbulence and Complex Systems, Peking University, Beijing, China, 100871}
\SetAffiliation{3}{Mechanical Engineering and Applied Mechanics, The University of Pennsylvania, Philadelphia, Pennsylvania, 19104}
\SetAffiliation{4}{The State Key Lab of Nonlinear Mechanics, Institute of Mechanics, Chinese Academy of Sciences, Beijing, China, 100190}


\maketitle





\begin{abstract}
We examine and benchmark the emerging idea of applying the large-eddy simulation (LES) formalism to unconventionally coarse grids where RANS would be considered more appropriate at first glance. 
We distinguish this idea from very-large-eddy-simulation (VLES) and detached-eddy-simulation (DES), which require switching between RANS and LES formalism.
LES on RANS grid is appealing because first, it requires minimal changes to a production code; second, it is more cost-effective than LES; third, it converges to LES; and most importantly, it accurately predicts flows with separation. 
This work quantifies the benefit of LES on RANS-like grids as compared to RANS on the same grids.
Three canonical cases are considered: periodic hill, backward-facing step, and jet in cross flow. 
We conduct direct numerical simulation (DNS), proper LES on LES grids, LES on RANS-quality grids, and RANS. 
We show that while the LES solutions on the RANS-quality grids are not grid converged, they are twice as accurate as the RANS on the same grids.
\end{abstract}


\begin{nomenclature}
\EntryHeading{Roman letters}
\entry{$L$}{domain size}
\entry{$R_{ij}$}{Reynolds stress}
\entry{$S_{ij}$}{strain rate tensor}
\entry{$u_i$}{instantaneous velocity vector}
\entry{$x,y,z$}{three Cartesian coordinates}

\EntryHeading{Greek letters}
\entry{$\Delta x,y,z$}{grid spacing in the three directions}
\entry{$\eta$}{Kolmogorov scale}
\entry{$\nu$}{kinematic viscosity}

\EntryHeading{Dimensionless groups}
\entry{Re}{Reynolds number}

\end{nomenclature}


\section{Introduction}
\label{sect:intro}

Reynolds-averaged Navier Stokes (RANS) has been used extensively in engineering since the 1990s.  
It solves for the mean flow directly while the entirety of turbulence is modeled \cite{pope2000turbulent}, and therefore is not a scale-resolving tool.
Large-eddy simulation (LES) is a scale resolving tool, and it is seeing more use in fluids engineering \cite{goc2021largeJ,zamiri2020large,cheng2021large,zhao2019large}.
Nonetheless, RANS is the go-to tool for engineering design work. 

In general, scale-resolving techniques are more accurate than non-scale-resolving ones \cite{slotnick2014cfd}.
However, the above statement applies to properly grid-converged results obtained with the two respective approaches.
Grid-converged results is not always obtainable with scale-resolving tools, especially at high Reynolds number applications, due to the their high computational cost \cite{choi2012grid,yang2021grid}.
Here, we briefly review the convergence trends in RANS and LES. 
The turbulence models deployed in RANS are designed and meant to be responsive to variations at the scales of the mean flow only. 
They are diffusive, tending to suppress the generation of scales smaller than the mean flow scales. 
As a result, the grid convergence in RANS takes place much earlier compared to LES. 
This converged solution would correspond to the exact solution of the model equations rather than the Navier-Stokes equations.
Consequently, model errors in RANS do not diminish with (or respond to) grid refinements, in contrast to LES subgrid models, which converge to the Navier Stokes equation with sufficiently fine grid resolution.  

Recent LES-based studies of high-Reynolds number external aerodynamics configurations produced interesting observations in this context. 
Park and Moin employed a very coarse grid when computing a flow over the NASA common research model (CRM) with wall-modeled LES \cite{park2016wall}. 
The grid used had  {$12\times10^6$} locally isotropic cells, resolving the wing boundary layer on average with up to 5 cells only. 
The grid used was more a RANS grid than a proper LES grid, and this calculation was not grid-converged. 
Despite these apparent inadequacies, the lift and drag forces were predicted with reasonable accuracy. 
{Goc et al. \cite{goc2021largeJ}} made similar observations in their wall-modeled LES of flow over the JAXA Stanford Model (JSM) in a high-lift configuration. 
They were able to get a reasonably accurate lift and drag using  {$9\times10^6$} grid cells at AOA from 5 to 22 degrees (at pre and post stall conditions). 
Goc et al. further noted that coarse grid LESs give more accurate results than RANSs on the same coarse grid \cite{goc2021large}. 
These calculations can effectively be viewed as LES calculations conducted on grids that are coarse enough to be more suitable for RANS but produce better results than the usual RANS. 
This brings us to the topic of this work, LES on RANS-quality grid, the concept of which was discussed earlier from the industrial perspective in {Ref.} \cite{xu2021comparative}, where the purpose was to make use of existing RANS grids that no longer offer improvements in RANS predictions with further grid refinement.

From a practical point of view, computational affordability and short turnaround time is as important as accuracy. 
The trade-off between the two heavily affects the choice between LES and RANS. 
Take the NASA CRM as an example: the size of a typical RANS grid is O(10) millions \cite{park2016wall,rumsey2020reynolds},  and the size of a proper (wall-modeled) LES grid is O(1) billion \cite{park2016wall, goc2021largeJ}. 
For an engineer with access to a few hundred CPU cores, an O(10) million grid-point calculation is quite affordable, but an O(1) billion grid-point calculation is not.
On the other hand, if one sticks with RANS, the results stop improving before exhausting the computing resources, but if proper LES is desired, there are not enough computational resources.
The practical question is: given an affordable grid that does not offer further improvements from RANS with mesh refinements, would it be better to solve LES equations on this grid?
This concept of LES on a RANS-quality grid in a cost-driven industry is attractive. 
The computational costs of a RANS and an LES on a RANS-quality grid are comparable---LES on a RANS-quality grid would be slightly more costly since the instantaneous flow fields require time averaging.

The purpose of this work is to benchmark this idea of LES on a RANS-quality grid.
The benchmark here focuses on separated flows, which remains a challenge to predictive CFD \cite{witherden2017future}. 
Three canonical cases are considered, namely, periodic hill, backward-facing step, and jet in cross flow.
Flow separates as a result of the surface geometry in the first two cases and as a result of cross flows in the third case.
The geometry of the periodic hill gives rise to smooth-body separation, and the geometry of the backward-facing step gives rise to bluff-body separation.
Among the three cases, the periodic hill case and the backward-facing step case are widely used for RANS and LES validation and verification \cite{zhou2021reynolds, xiao2020flows,chaouat2013hybrid,hanjalic1998contribution,de2018use}.
We conduct direct numerical simulations (DNSs), LESs, LESs on RANS-quality grids, and RANSs.
The simulation details are presented in Section \ref{sect:grid}.
An emphasis is put on grid construction.
The simulation results are presented in Section \ref{sect:results}, and we provide a explanation of why LES on RANS-quality grids are successful in Section \ref{sect:discussion}.
Lastly, we conclude in Section \ref{sect:conclusions}.

\section{Computational details}\label{sect:grid}

\subsection{Principles for meshing and complications}
\label{sub:2.1}

\begin{figure}
\centering
\includegraphics[width=0.4\textwidth]{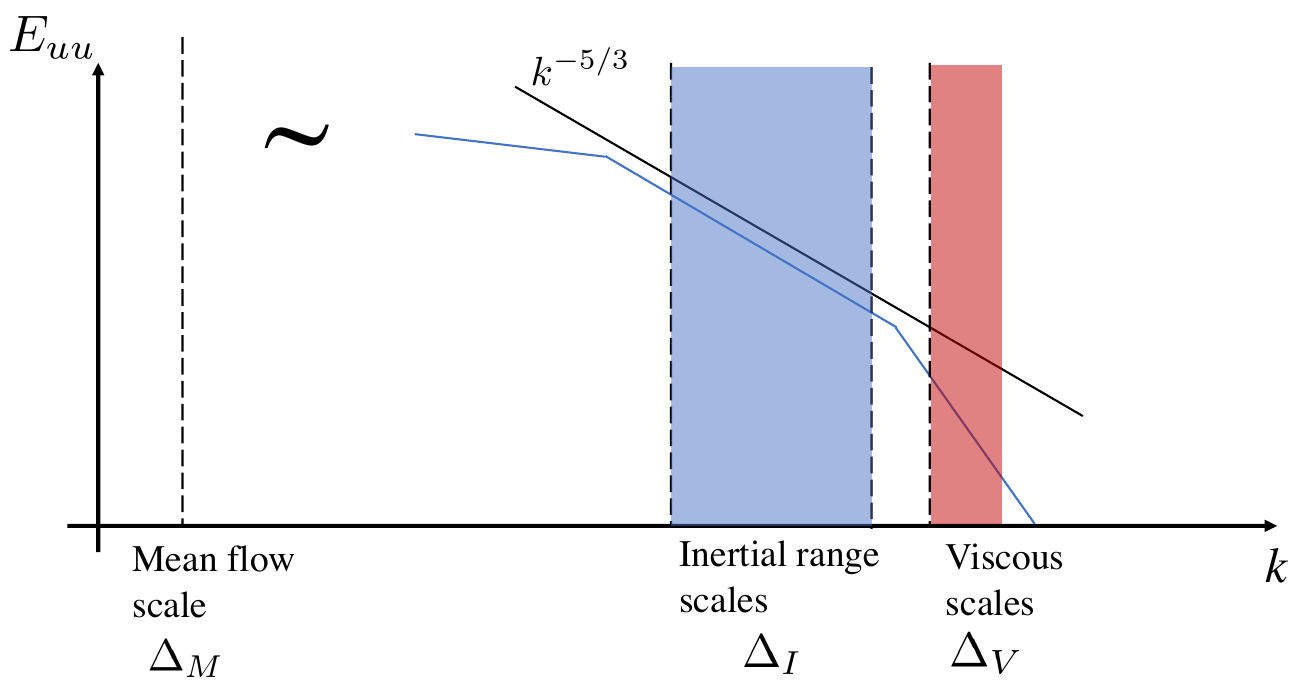}
\caption{A sketch of a generic turbulent energy spectrum and the various length scales in a turbulent flow.
At sufficiently high Reynolds numbers, the mean flow scale, the inertial range scale(s), and the viscous scale(s) are well separated.
}
\label{fig:scales}
\end{figure}

We must define DNS grids, LES grids, and RANS grids.
Following the discussion in Ref. \cite{pope2000turbulent}, 
we review the basic grid resolution requirements for RANS, LES, and DNS.
A RANS grid should resolve the mean flow and its gradients, requiring the grid spacing to scale with the mean flow scales $\Delta_M$.
An LES grid should resolve the large-scale turbulent eddies.
Ideally, the grid spacing $\Delta_I$ should be in the inertial range.
A DNS grid should resolve the viscous scales, and ideally, the grid spacing $\Delta_V$ should scale with the Kolmogorov length scale (2 to 4 Kolmogorov length scales).
Figure \ref{fig:scales} is a sketch of a generic turbulent energy spectrum, and we have also sketched the mean flow scale $\Delta_M$, the inertial range scale $\Delta_I$, and the viscous scale, $\Delta_V$.
If the Reynolds number is sufficiently high, we have:
First, the mean flow scale is much larger than any turbulence length scale.
Second, turbulence is roughly isotropic at sufficiently small scales.
Third, the inertial range scale is much larger than the Kolmogorov length scale.
It follows that, at sufficiently high Reynolds numbers, a RANS grid is much coarser than an LES grid, and an LES grid is much coarser than a DNS grid.

Applying these gridding principles to real-world applications is not always straightforward. 
Firstly, we usually do not know $\Delta_M$, $\Delta_I$, $\Delta_V$ {a priori}.
Secondly, all real-world flows are at finite Reynolds numbers, and the three scales are not always well separated. 
For example, complex geometries give rise to mean flow scales that are spatially varying. 
These mean flow scales are comparable to turbulence scales in places where the local Reynolds number is not high. 
In this situation, the grid spacing in RANS would be comparable to that in LES, as in  Ref. \cite{hanson2019flow}.
Thirdly, empiricism plays an important role.
For example, the viscous sublayer is well-resolved if the grid is such that $\Delta x^+\times \Delta y^+\times \Delta z^+\approx 12\times 0.05\times 6$.
Here, $\Delta x$, $\Delta y$, and $\Delta z$ are the grid spacings in the streamwise, wall-normal, and spanwise directions, and the superscript $+$ denotes normalization by the viscous wall units.
The Kolmogorov length scale is $\eta^+\approx 1$ at the wall. 
The grid spacing in the streamwise and the spanwise directions, i.e., $\Delta x^+=12$ and $\Delta z^+=6$, do not resolve the Kolmogorov length scale.
Nonetheless, $\Delta x^+=12$, $\Delta z^+=6$ still constitutes a DNS grid, which is an empiricism that does not conform to the general principle.
Fourth, the grid and the turbulence model interact, leading to strong sensitivity of critical flow events to the grid content. 
The grid-caused separation is such an example, {as Ref.\cite{menter2003ten}}, where slight changes in the grid may result in large changes in the resulting RANS solution.

Considering these real-world complications, we choose to design the grids to be used for the present study carefully (instead of using openly available existing grids) so that we could stay as close to the basic gridding principles as possible in arriving at a general conclusion about LES on RANS-quality grids.

\subsection{Meshing strategy}
\label{sub:2.2}

\begin{figure*}
\centering
\includegraphics[width=0.8\textwidth]{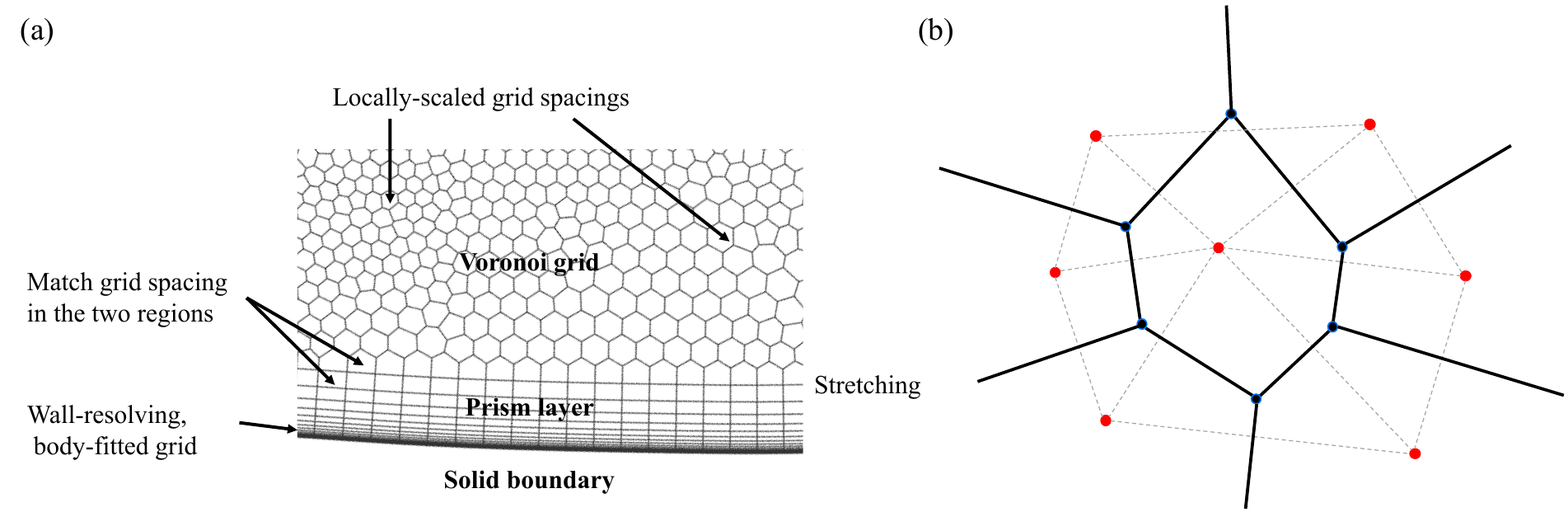}
\caption{(a) A sketch of the general meshing strategy.
Prism layers are employed near solid boundaries.
The viscous sublayer is resolved.
The wall-normal grid spacing is stretched until the grid aspect ratio is about unit.
A Voronoi grid is used away from solid boundaries, where the grid spacing scales locally per the requirement of DNS, LES, and RANS.
(b) A sketch of a Voronoi grid.
The grid is the median dual mesh of the primary (dashed) triangular mesh.
}
\label{fig:grid}
\end{figure*}

Figure \ref{fig:grid} (a) shows the general meshing strategy.
The grid is a Voronoi grid at a distance from solid boundaries.
Voronoi grids are generated from point clouds, as shown in Fig.\ref{fig:grid} (b).
By controlling the clustering of these seeding points, one can precisely control the grid spacing.
For instance, the resulting grid will be a RANS, LES, or DNS grid if we cluster the points such that their local density is $1/\Delta_M^3$, $1/\Delta_I^3$, and $1/\Delta_V^3$.
In addition to allowing us to precisely control the grid spacing, a Voronoi grid enables more accurate central reconstruction and better numerical stability than, say, triangular unstructured meshes \cite{fortune1995voronoi}.
The reader is directed to {Ref.} \cite{bres2018large,goc2021large,lozano2021performance} for further details about Voronoi grids and its recent applications.
A different strategy is employed near the solid boundaries, where we place anisotropic prism layers.
The wall-normal grid spacing is small at the wall and increases away from the wall until the grid aspect ratio is about unity, and the grid transitions smoothly to a locally isotropic Voronoi grid.
The exact transition location varies from one grid to another but usually is at about 20\% of the local boundary layer thickness. 
{In our work, a Delaunay-based mesh-generation code is employed \cite{engwirda2015voronoi,engwirda2018generalised} to generate the point cloud and a triangular mesh. 
The Voronoi mesh is the median dual mesh of the triangular mesh.}

\subsection{Equations}
\label{sub:2.2.2}

In DNS, the incompressible Navier-Stokes and continuity equations are solved 
on a sufficiently fine grid without any modeled terms:
\begin{eqnarray}
    \frac{\p u_j}{\p x_j} &=& 0,\\
    \frac{\p u_i}{\p t} + \frac{\p }{\p x_j}\left( u_i u_j \right) &=& -\frac{1}{\rho}\frac{\p p}{\p x_i} + \nu \frac{\p^2 u_i}{\p x_j^2},
    \label{eq:mom-dns}
\end{eqnarray}
where $u_i$ is the velocity in $i$th Cartesian direction, $\rho$ is the fluid density, $p$ is the pressure, and $\nu$ is the kinematic viscosity.

In LES, the following  filtered conservation equations are solved  on a grid that resolves part of the turbulence scales
\begin{eqnarray}
    \frac{\p \bar{u}_j}{\p x_j} &=& 0,\\
    \frac{\p \bar{u}_i}{\p t} + \frac{\p }{\p x_j} \left( \bar{u}_i \bar{u}_j \right)&=& -\frac{1}{\rho}\frac{\p \bar{p}}{\p x_i} + \nu \frac{\p^2 \bar{u}_i}{\p x_j^2} - \frac{\p \tau_{ij}}{\p x_j},
    \label{eq:mom-les}
\end{eqnarray}
where $\bar{\cdot}$ denotes a filtered quantity,
$\tau_{ij} \equiv \overline{u_iu_j} - \bar{u}_i\bar{u}_j$ is the subgrid-scale (SGS) stress. 
Invoking the eddy-viscosity model, the anisotropic part of the SGS stress is modeled as
\begin{equation}
\small
    \tau_{ij} - \frac{1}{3}\tau_{kk}\delta_{ij} = -2\nu_{sgs}\bar{S}_{ij},
\end{equation}
where $\delta_{ij}$ is the Kronecker delta, $\bar{S}_{ij} \equiv (\p \bar{u}_i/\p x_j + \p \bar{u}_j/\p x_i)/2$ is the filtered rate-of-strain tensor, $\nu_{sgs}$ is the SGS eddy viscosity modeled with the WALE (Wall-Adapting Local Eddy-viscosity) model. 
The reader is directed to {Ref.} \cite{nicoud1999subgrid} for the details of the WALE model.

Last, the following Reynolds-averaged equations are solved  on a grid that resolves the mean flow
\begin{equation}
\small
\centering
\begin{split}
\small
    \frac{\p \<u_j\>}{\p x_j} &= 0,\\
    \frac{\p \<u_i\>}{\p t} + \frac{\p }{\p x_j} \left( \<u_i\> \<u_j\> \right) &= -\frac{1}{\rho}\frac{\p \<p\>}{\p x_i} + \nu \frac{\p^2 \<u_i\>}{\p x_j^2} - \frac{\p R_{ij}}{\p x_j}.
\end{split}
\label{eq:mom-rans}
\end{equation}
where $\<\cdot\>$ denotes Reynolds averaging,
$R_{ij}$ is the Reynolds stress and, here, is modeled according to 
\begin{equation}
\small
    R_{ij} - \frac{1}{3}R_{kk}\delta_{ij} = -2\nu_t\<S_{ij}\>,
\end{equation}
where $\nu_t$ is the eddy viscosity and is modeled per the $k$-$\omega$ SST model \cite{menter1994two} or the SA model \cite{spalart1992one}, and $\<S_{ij}\>$ is mean strain rate tensor.

\subsection{Benchmark cases: computational setup}
\label{sub:2.4}

\begin{figure}
  \centering
  \includegraphics[width=0.45\textwidth]{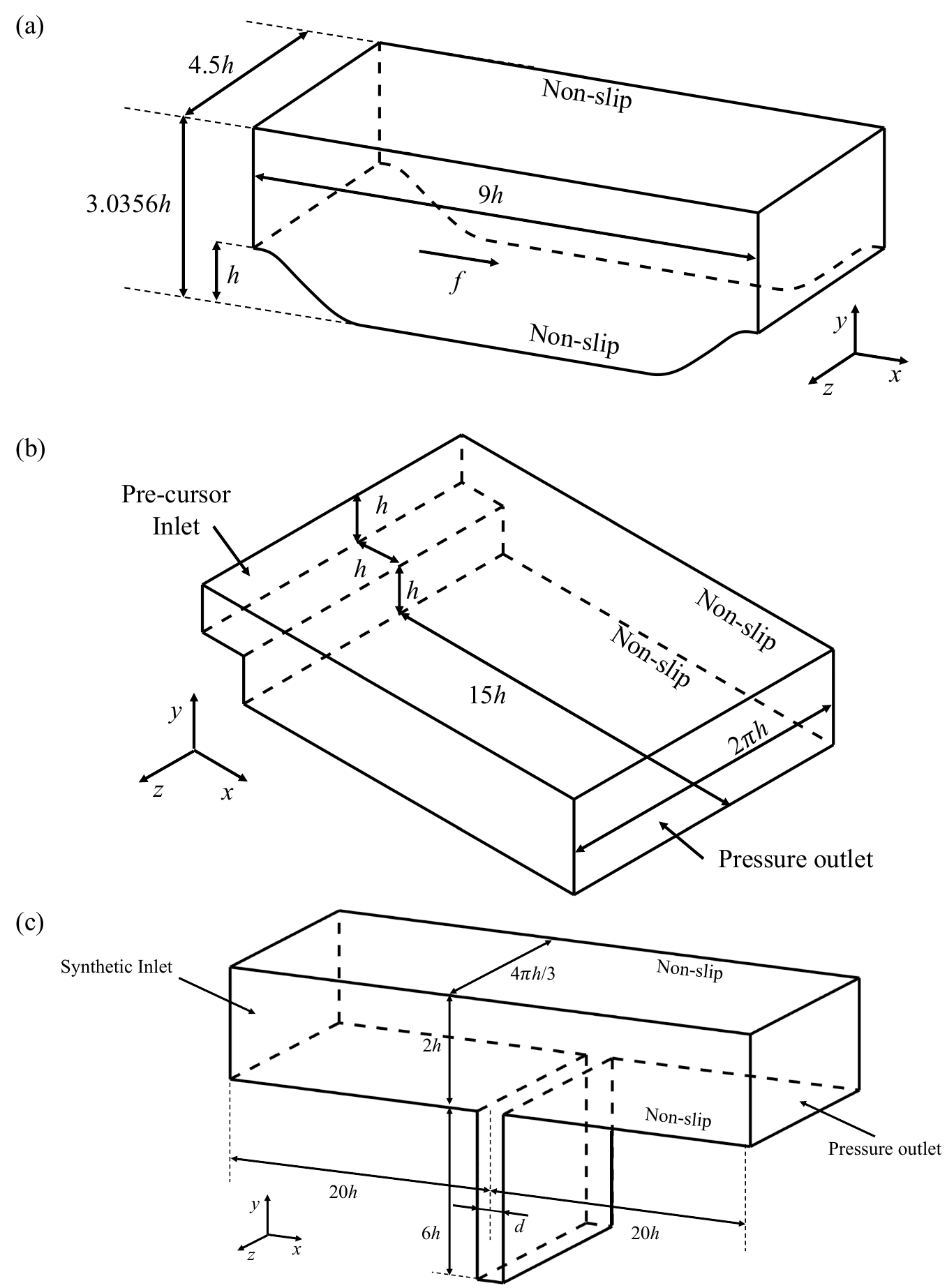}
  \caption{Schematics of (a) periodic hill (b) backward-facing step, and (c) jet in cross flow. \label{fig:sketch-case}}
\end{figure}

The cases considered here are periodic hill, backward-facing step, and jet in cross flow. 
The geometries, boundary conditions, and grids used are described first in the forthcoming sections. 
The simulation results will be presented separately in section \ref{sect:results}. 

Figure \ref{fig:sketch-case} (a) shows a schematic of the periodic hill, 
the geometry of which is described in detail in Ref. \cite{zhou2021wall}. 
The size of the domain is $L_x\times L_y\times L_z=9h\times 3.0356h\times 4.5h$, where $h$ is the height of the hill. 
The flow is doubly periodic in the streamwise ($x$) and the spanwise ($z$) directions, while it is statistically homogeneous only in the $z$ direction.  
The top and bottom boundaries are the no-slip walls. 
The flow is driven by a body force in the $x$ direction.
The bulk Reynolds number is $Re_{b}=U_b h/\nu=5600$, where $U_b$ is the bulk velocity, and $\nu$ is the kinematic viscosity.
We will consider a higher Reynolds number in Section \ref{sect:discussion}.

\begin{figure}[htb!]
  \centering
  \includegraphics[width=0.24\textwidth]{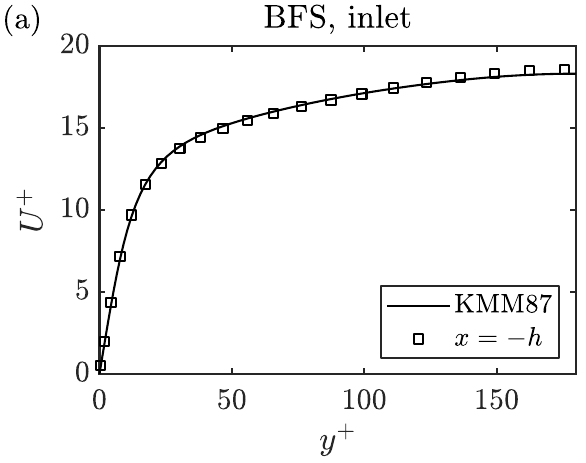}\includegraphics[width=0.24\textwidth]{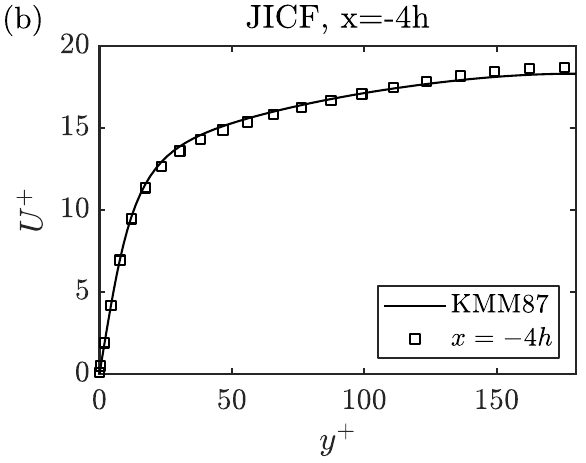}
  \caption{\label{fig:sjUMeanInlet} The velocity profiles (a) at the inlet of the back facing step case and (b) at $x=-4h$ of the jet-in-cross-flow case. \label{fig:UMeanInlet}}
\end{figure}

Figure \ref{fig:sketch-case} (b) is a schematic of the backward-facing step.
The step is located at $x=0$, and its height is $h$.
The size of the domain is $L_x\times L_y\times L_z=16h\times 2h\times 2\pi h$.
The computational inlet is located at $x=-h$ and the outlet at $x=15h$.
The instantaneous fluctuating inflow condition required in LES and DNS is 
extracted from a separate precursor simulation of 
a fully developed channel flow at  $Re_\tau=180$.
Figure \ref{fig:UMeanInlet} (a) shows the mean velocity profile at $x=-h$.
{The pressure is fixed at the outlet, and the zero Neumann condition is imposed for the velocity at the outlet. The pressure at the inlet is computed instead of imposed.}
Periodicity is imposed in the $z$ direction.
Both the bottom and top boundaries are no-slip walls.
The bulk Reynolds number is $Re_b=U_bh/\nu=5641$.

Figure \ref{fig:sketch-case} (c) is a schematic of the jet-in-cross-flow case.
A fully developed turbulent channel encounters a 2D vertical plane jet at $x=0$.  
The size of the channel is $L_x\times L_y\times L_z=40h \times 2h \times 4\pi h/3$, and the size of the feed is $L_x'\times L_y'\times L_z'=d\times 6h\times 4\pi h/3$, where $d=0.5h$.
The inlet is located at $x=-20h$, and the outlet at $x=20h$.
A fully developed channel flow is prescribed at the inlet ($Re_\tau=180$). 
Fluctuating inflow turbulence is synthesized at the inlet when needed (in LES and DNS) using the approach in {Ref.} \cite{xie2008efficient}.

\begin{figure}
\centering
\includegraphics[width=0.47\textwidth]{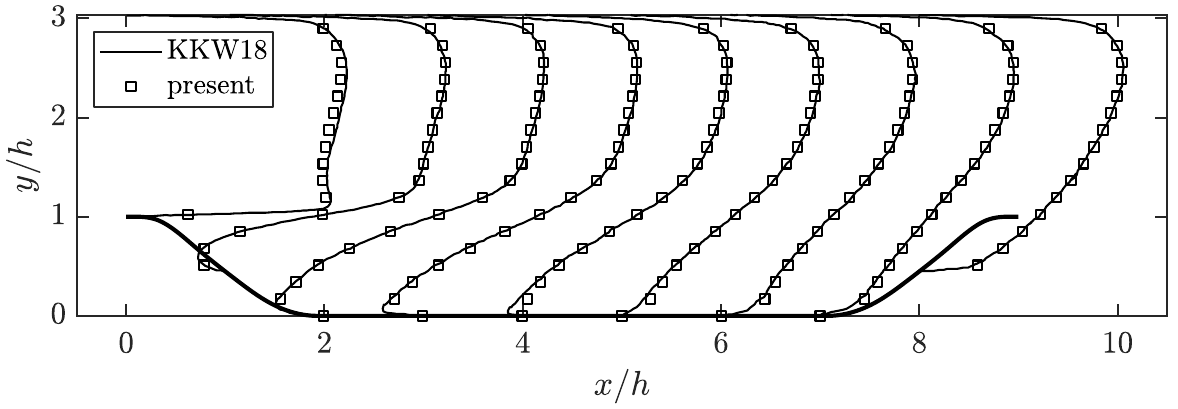}
\caption{Velocity profiles at $x/h = 0$, 1, 2, 3, 4, 5, 6, 7 and $8$. We compare our DNS and that in \cite{krank2018direct} (KKW18 in the figure).}
\label{fig:phSketch}
\end{figure}

\begin{figure}
    \centering
    \includegraphics[width=0.47\textwidth]{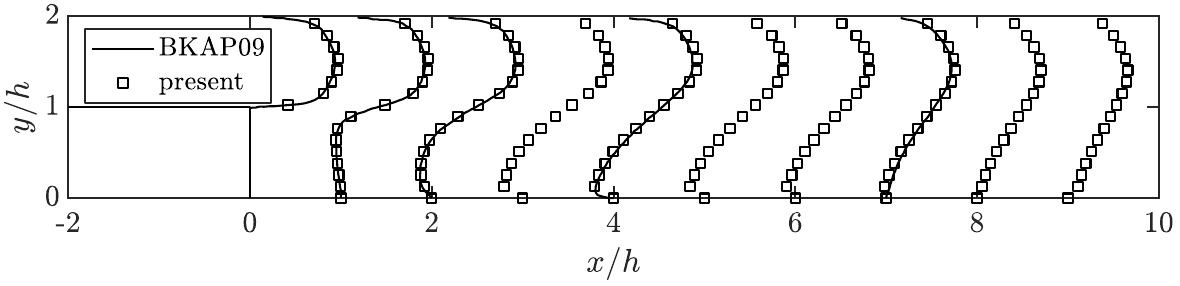}
    \caption{Velocity profiles at $x/h=0, 1, 2, 3, 4, 5, 6, 7, 8,$ and $9$. We compare our DNS and that in \cite{barri2009dns} (BKAP09 in the figure).}
\label{fig:bfsUmean}
\end{figure}

Figure \ref{fig:sjUMeanInlet} {(b)} shows the mean velocity profile at $x=-4h$.
The jet velocity is $U_j=U_b$ at $y=-6h$, corresponding to a blowing ratio of 1.
Here, $U_j$ is the jet velocity, and $U_b$ is the bulk velocity of the incoming channel flow.

\subsection{Benchmark cases: DNS results}
\label{sub:2.5}

We present the DNS results of the  benchmark cases here.
The purposes of the DNSs are first to validate our code, second to obtain the Kolmogorov length scale distribution to guide the grid generation, and third to serve as the reference data for LESs, RANSs, and LESs on RANS-quality grids. 
We would scale the LES and RANS grid spacings such that they are multiples of the local Kolmogorov length scale, $\eta=(\nu^3/\epsilon)^{0.25}$, where $\epsilon$ is the dissipation rate and is available only in DNSs.
Here, we follow {Ref.} \cite{krank2018direct,barri2009dns} and repeat their DNSs.
DNS of jet in cross flow can be found in Xu et al. \cite{xu2022direct}.

Figure \ref{fig:phSketch} shows the velocity profiles at 9 streamwise locations from the periodic hill calculation. 
We compare our DNS result to that in {Ref.} \cite{krank2018direct}, where a very good agreement is found. 
Figure \ref{fig:bfsUmean} shows the velocity at 10 streamwise locations downstream of the backward-facing step, which is compared to the DNS in {Ref.} \cite{barri2009dns} and shows excellent agreement.

\subsection{Benchmark cases: grids}
\label{sub:2.6}

We scale the Voronoi grid spacing such that they are 7.5, 15, and 25 times the local Kolmogorov length scale.
{These grids are generated by employing the meshing strategy outlined in Section \ref{sub:2.2}.}
We refer to the three grids as the L-grid, the LR-grid, and R-grid, where the L-grid is LES-quality, and the R-grid is a fairly coarse RANS grid.
{RANS and LES equations are solved on all three grids.
LES on the L-grid is a proper LES.
LES on the R-grid is LES on RANS-quality grid.}
LES on the RANS-quality grid does not require the grid spacing to be proportional to the local Kolmogorov length scale.
The purpose of scaling the grid spacing to the local Kolmogorov length scale is to systematically control the grid quality.

\begin{figure}[htb]
  \centering
  \includegraphics[width=0.47\textwidth]{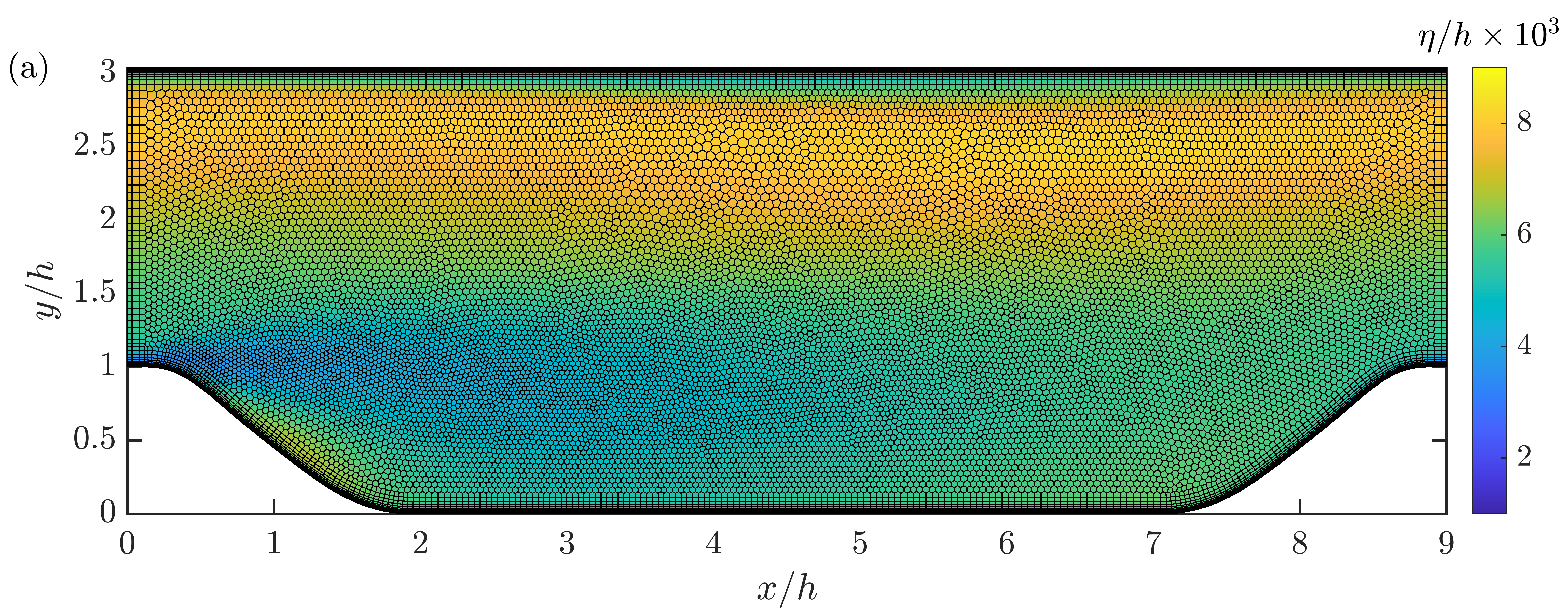}
  \includegraphics[width=0.47\textwidth]{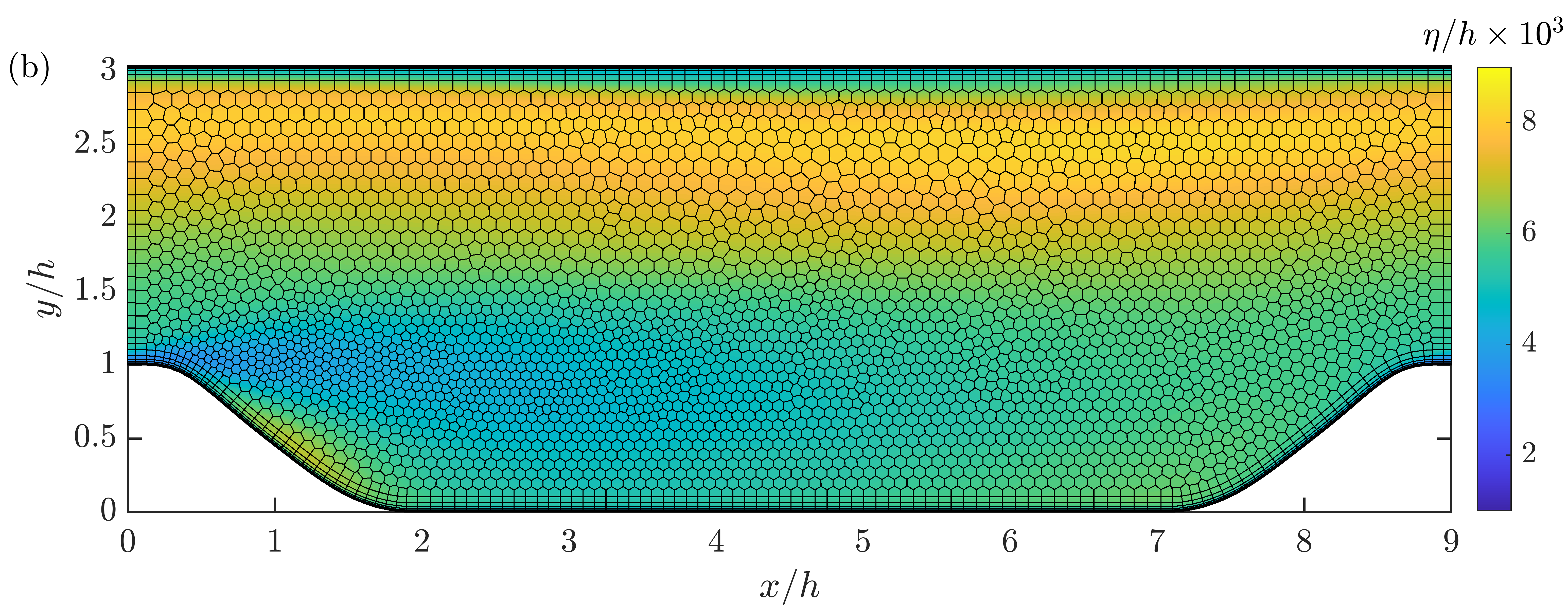}
  \includegraphics[width=0.47\textwidth]{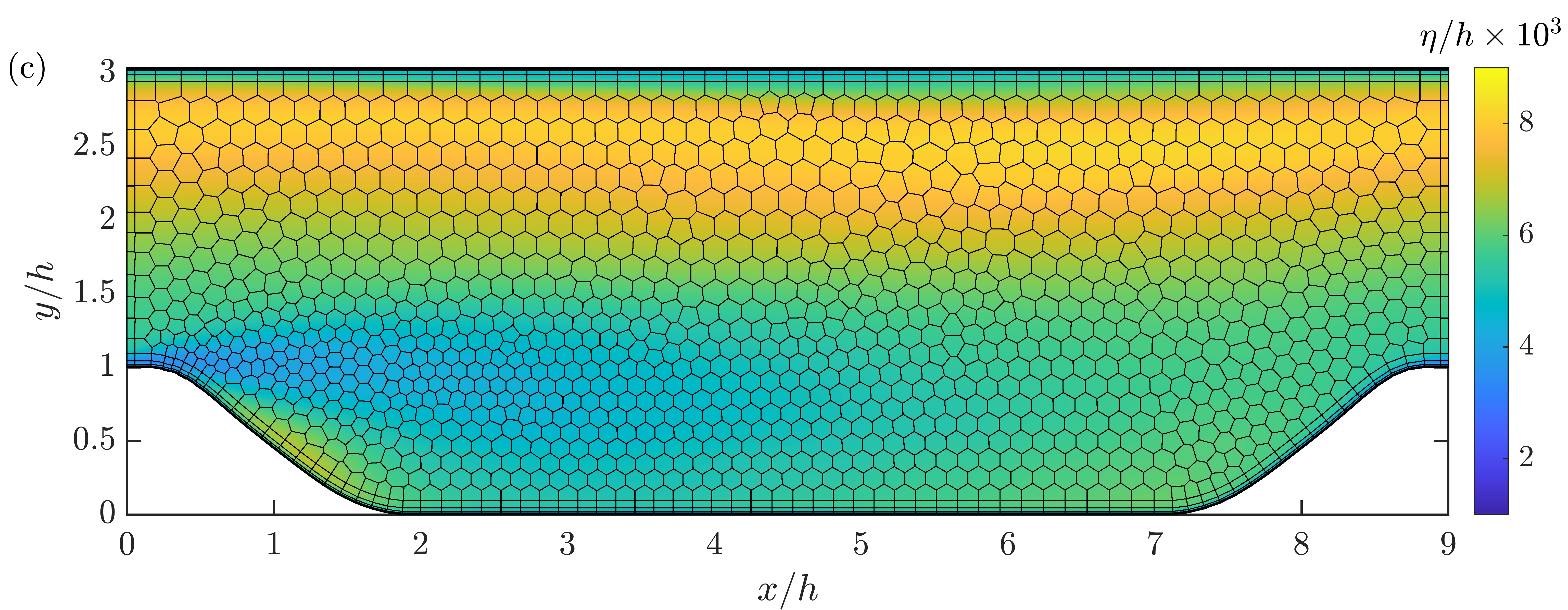}
  \caption{(a) L-grid, (b) LR-grid, and (c) R-grid for the periodic hill. 
  The contours show the Kolmogorov length scale computed from our DNS.}
  \label{fig:phMesh}
\end{figure}

Figure \ref{fig:phMesh} shows the L-grid, LR-grid, and R-grid for the periodic hill.
Figure \ref{fig:phMesh} also shows the contours of the local Kolmogorov length scale.
The presence of the hill gives rise to smaller turbulence scales in the bottom half of the channel than the top half of the channel, with the smallest scales found in the shear layer downstream of the hill.
The L-grid, LR-grid, and R-grid have about {$21\times10^3$, $6\times10^3$, and $2\times10^3$} grid points in the x-y plane.
We compare these numbers to the ones in the literature: the LES grid in {Ref.} \cite{balakumar2014dns} has {$39\times10^3$} grid points in the plane, and the RANS grid in Ref. \cite{xiao2012consistent} contains about  {$3\times10^3$} grid points in the plane.

\begin{figure}
  \centering
  \includegraphics[width=0.47\textwidth]{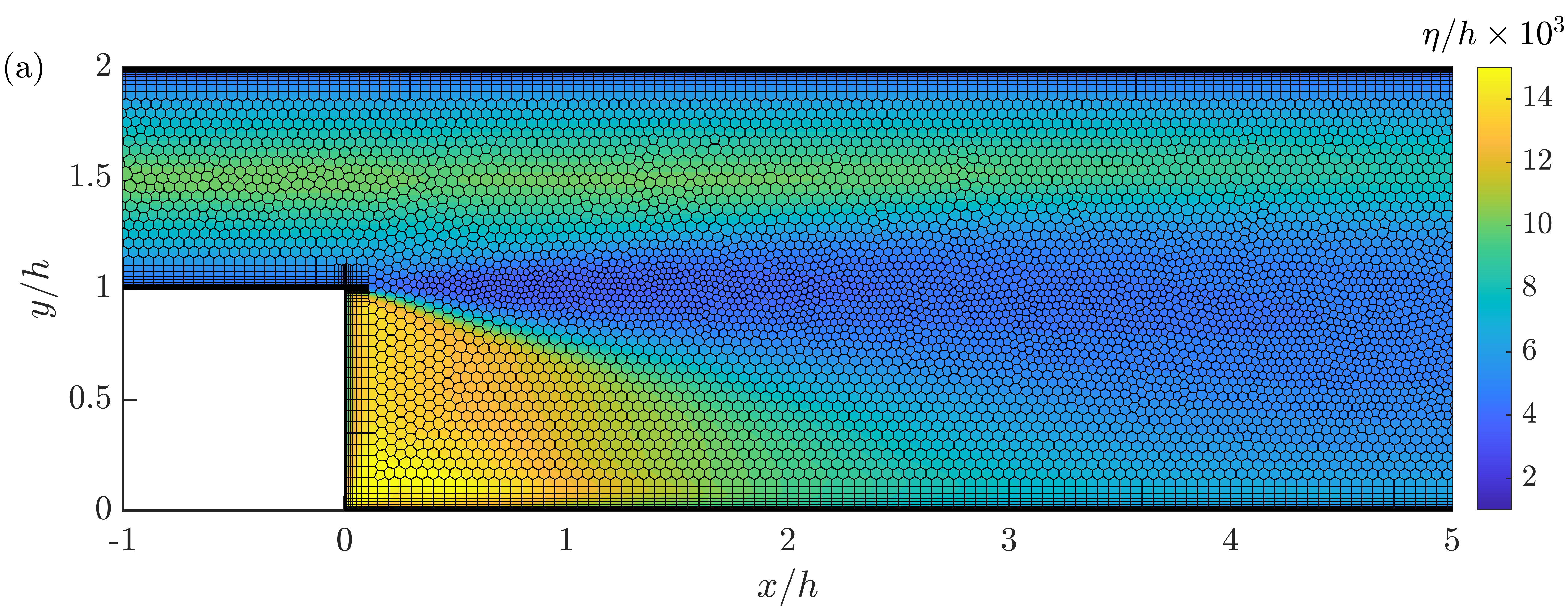}
  \includegraphics[width=0.47\textwidth]{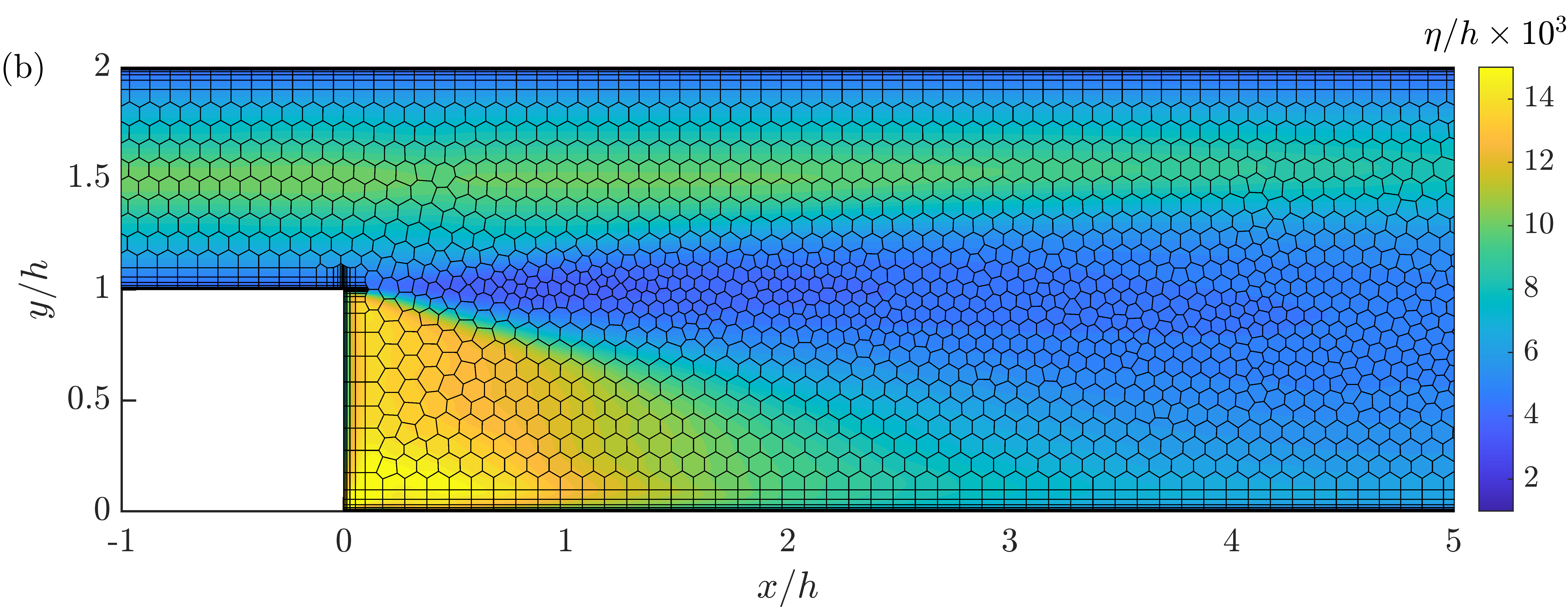}
  \includegraphics[width=0.47\textwidth]{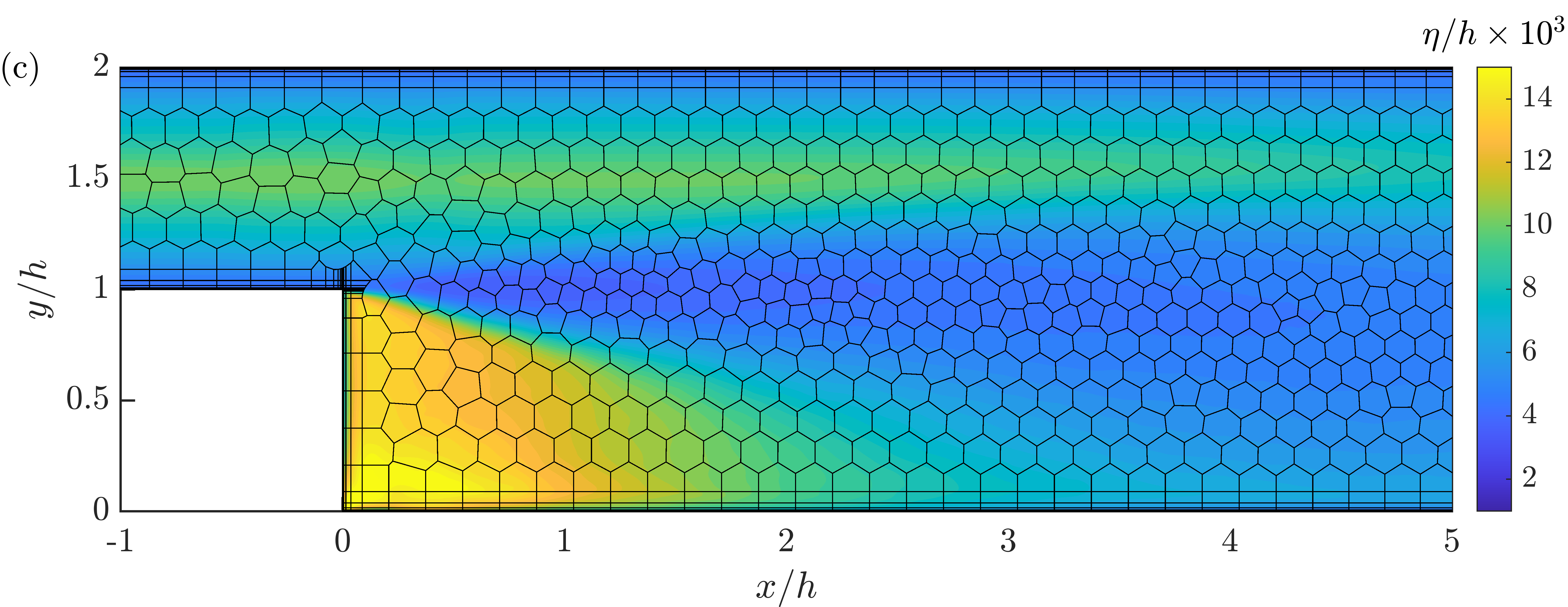}
  \caption{Same as Fig.\ref{fig:phMesh}, but for the back facing step. 
  }
  \label{fig:bfsMesh}
\end{figure}

\begin{figure}
  \centering
    \includegraphics[width=0.47\textwidth]{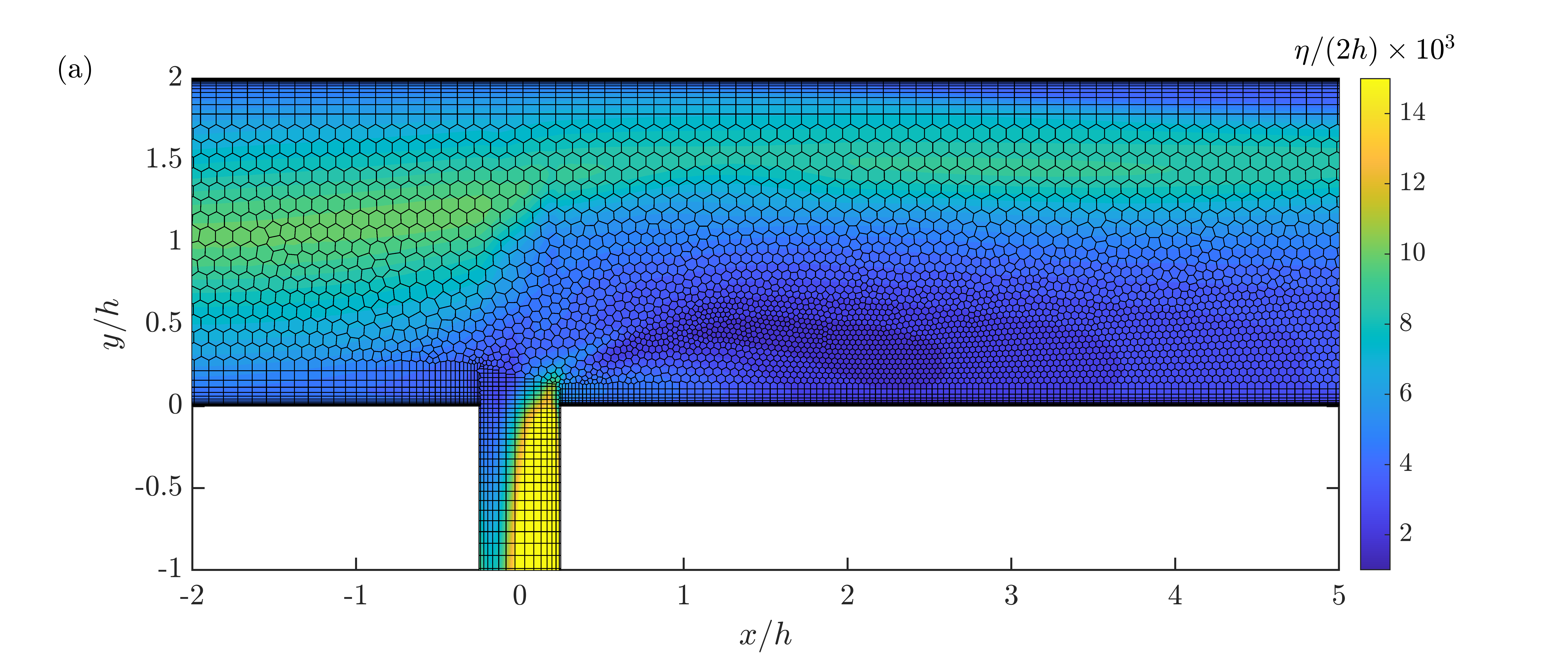}
    \includegraphics[width=0.47\textwidth]{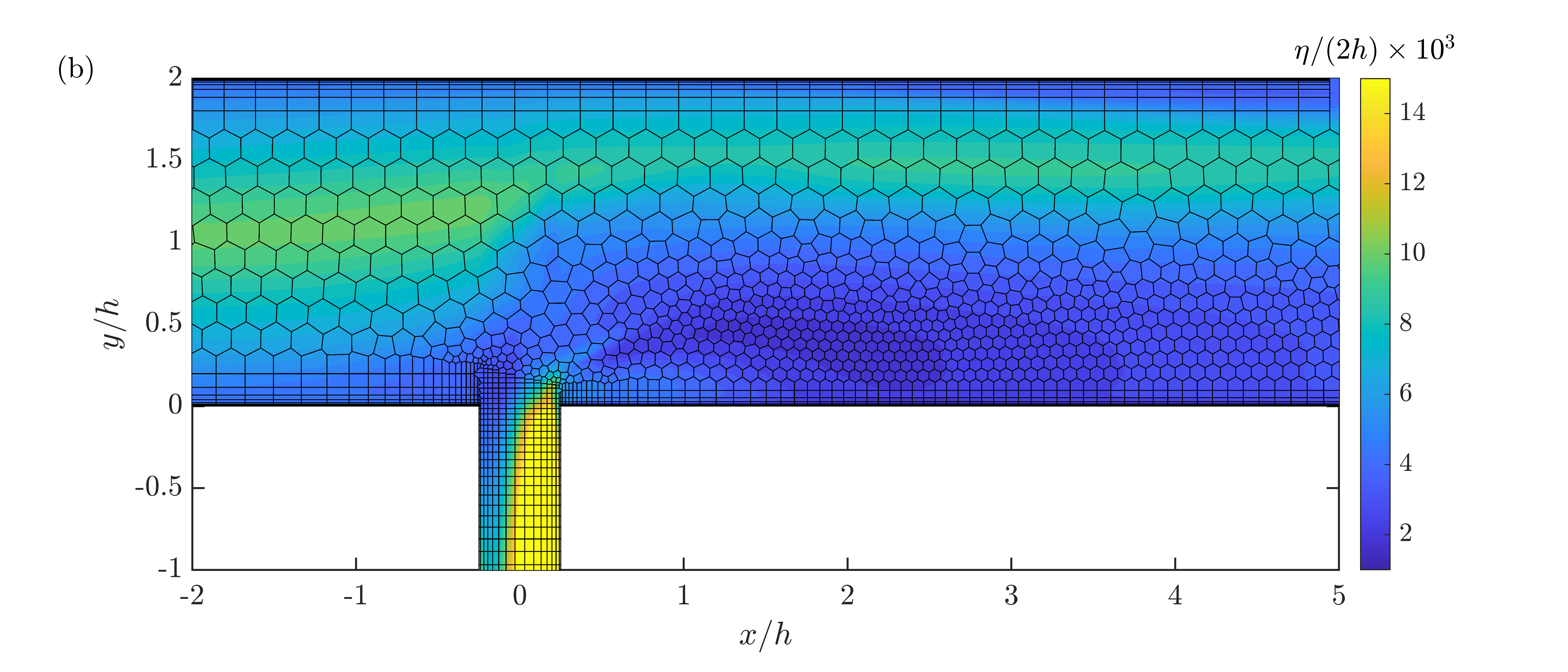}
    \includegraphics[width=0.47\textwidth]{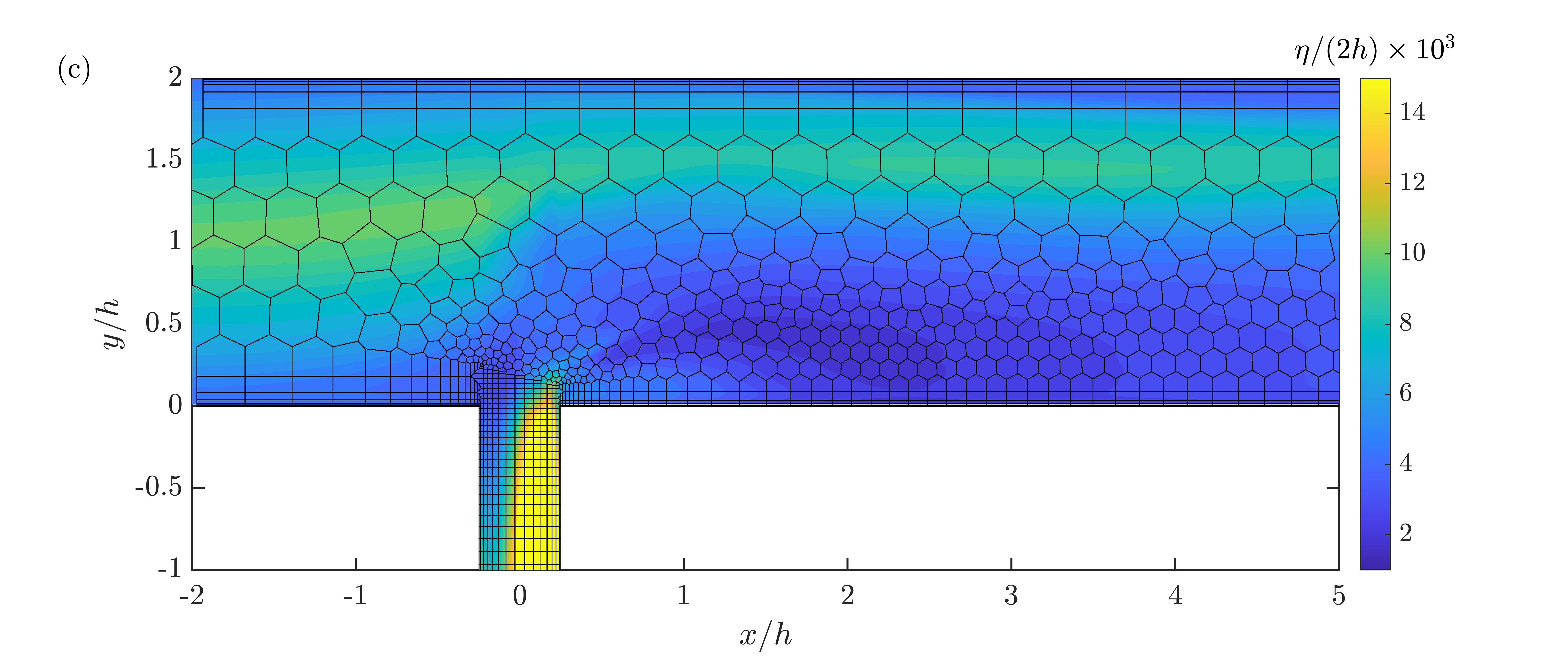}
  \caption{Same as Fig.\ref{fig:phMesh} but for jet in cross flow.}
  \label{fig:sjMesh}
\end{figure}

Figure \ref{fig:bfsMesh} shows the L-grid, LR-grid, and R-grid for the backward-facing step and contours of the local Kolmogorov length scale (made available by our DNSs).
The smallest turbulence scales are found in the shear layer near $0<x<h$, $y/h=1$.
Turbulence dissipation is small in the immediate wake of the step, leading to large turbulence scales there.
We require the grid spacing in the L-grid, LR-grid, and R-grid to be smaller than $0.05h$, $0.1h$, and $0.2h$.
These spacings translate to no fewer than 20, 10, and 5 grid points across the height of the $Re_\tau=180$ inlet channel.
The purpose of the limit is to prevent an overly under-resolved wake.
There are about  {$18\times10^3$, $5\times10^3$, and $2\times10^3$} grid points in the L-grid, LR-grid, and R-grid between $x=-h$ and $x=10h$ in the $x$-$y$ plane.
The numbers are  {$10\times10^3$} in {Juste et al.'s} \cite{juste2016assessment} LESs and {$6\times10^3$} in {Wu et al.'s} \cite{wu2013mixed} RANSs.
Again, the R-grid is a coarse RANS grid.

Figure \ref{fig:sjMesh} shows the L-grid, LR-grid, and R-grid for the jet in cross flow case and the Kolmogorov length scale.
The jet gives rise to strong turbulence downstream of its injection and small turbulence length scale near the bottom wall.
Again, we require the grid spacing to be smaller than 0.1$h$, 0.2$h$, and 0.4$h$ in the L-grid, LR-grid, and R-grid.
These grid spacings translate to 20, 10, and 5 grid points across the channel height.
The L-grid, LR-grid, and R-grid have  {$6.5\times10^3$, $2.5\times10^3$, and $1.6\times10^3$} grid points from $x=-h$ to $x=5h$.

Table \ref{tab:CaseDetail} shows the further details of the grids. 
The nomenclature is [Geometry][grid], where Geometry is PH, i.e., periodic hill, BFS, i.e., back facing step, or JCF, i.e., jet in cross flow, 
We list the approximate number of grids in the $x$-$y$ plane, the grid spacing, its cutoff, and the number of prism layers.

\begin{table}
\caption{\label{tab:CaseDetail} Further details of the grids.
Here, $n_c$ is the number of grid points in the x-y plane, $\Delta/\eta$ is the grid resolution, $\Delta_{\rm max}$ is the largest grid spacing (grid spacing cutoff), $h_{\rm prism}$ is the height of the prism layer, and $n_y$ is the number of grids across the prism layer. 
There are two numbers for the JCF case denoting the height of the prism layer upstream and downstream of the leakage jet.}
\small
\centering
\begin{tabular}{lcccccc}
\hline
Case & $n_c$&$\Delta/\eta$& $\Delta_{\rm max}$ & $h_{\rm prism}$ &$n_{y}$\\\hline
PH-L  & 20678& 7.5&     N/A  & $0.125h$            & 19\\ 
PH-LR &  5760&  15&     N/A  & $0.125h$            & 12\\
PH-R  &  2301&  25&     N/A  & $0.125h$            & 9\\
BFS-L & 25563& 7.5&  $0.05h$ & $0.125h$            & 16\\
BFS-LR& 7227&   15&   $0.1h$ & $0.125h$            & 11\\
BFS-R &  2936&  25&   $0.2h$ & $0.125h$            & 9\\
JCF-L & 25509& 7.5&   $0.1h$ & $0.25h$, $0.125h$   & 14\\
JCF-LR& 8136&   15&   $0.2h$ & $0.25h$, $0.125h$   & 10\\
JCF-R &  4038&  25&   $0.4h$ & $0.25h$, $0.125h$   & 7\\
\hline
\end{tabular}
\end{table}

\section{Results}\label{sect:results}

We present the periodic hill, backward-facing step, and jet in cross flow results in Secs. \ref{sec:ph}, \ref{sec:bfs}, and \ref{sec:sj}, respectively.
We examine velocity profiles at multiple streamwise locations, the skin friction and its error, with the latter more clearly showing differences among the solutions than the former.
We shall see that for a RANS grid that no longer offers much improvement for RANS with further grid refinement, solving the LES equations on that grid is beneficial, even though the LES on the RANS-quality grid is not grid converged.
Specifically, LESs on RANS-quality grids are considerably more accurate than RANS on RANS-quality grids in terms of their predictions of the skin friction coefficient.

\subsection{Periodic Hill}
\label{sec:ph}

\begin{figure}[htb]
    \centering
    \includegraphics[width=0.4\textwidth]{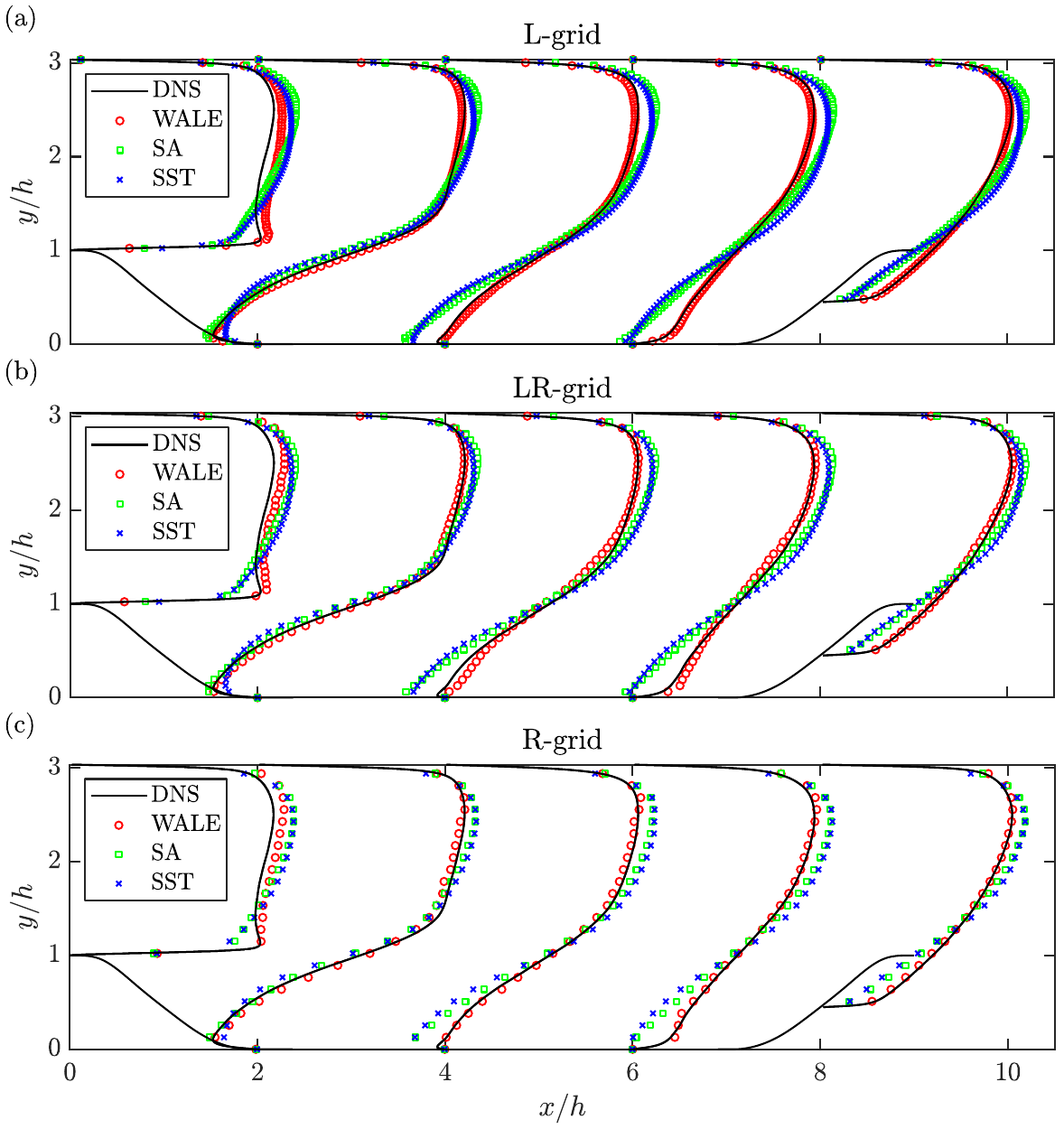}
    \caption{\label{fig:u-ph} Profiles of the $x$ direction velocity. (a) L-grid results. (b) LR-grid results. (c) R-grid results. }
\end{figure}

We compare the LES and RANS solutions on the three grids.
Figure \ref{fig:u-ph} shows the velocity profiles at several streamwise locations, and we compare the solutions on the L-grid, LR-grid, and R-grid.
We see that, for a given turbulence modeling approach, LES or RANS, the solutions do not differ across the L-grid, LR-grid, and R-grid.
The LES solution, i.e., WALE in Fig.\ref{fig:u-ph}, follows the DNS solution more closely than the two RANS solutions.

\begin{figure}[htb]
  \centering
  \includegraphics[width=0.24\textwidth]{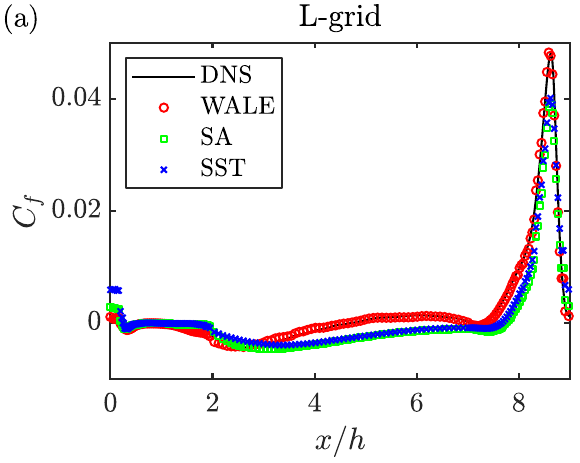}\includegraphics[width=0.24\textwidth]{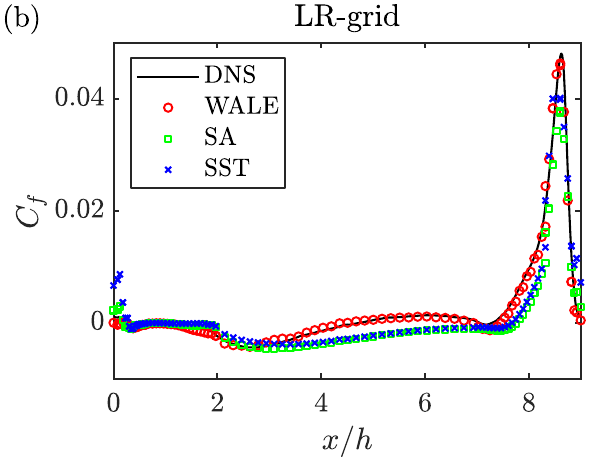}
  \includegraphics[width=0.24\textwidth]{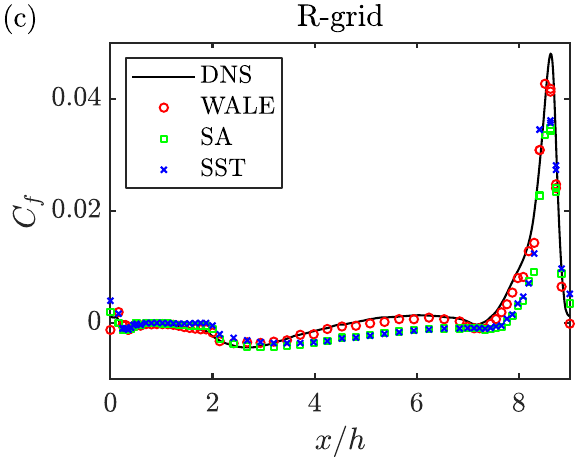}
  \caption{\label{fig:phCf}The skin friction coefficients on (a) the L-grid, (b) the LR-grid, and (c) the R-grid.}
\end{figure}

\begin{figure}[htb]
  \centering
  \includegraphics[width=0.27\textwidth]{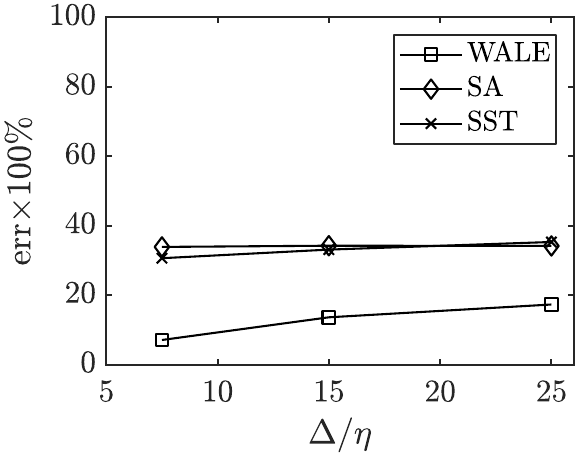}
  \caption{\label{fig:pherr} The errors in $C_f$ as measured by Eq. \ref{eq:pherr} as a function of the grid resolution. 
  The three data points on each line are the L-grid, LR-grid, and R-grid results, respectively, from left to right.}
\end{figure}

Figure \ref{fig:phCf} shows the skin friction coefficient $C_f$
\begin{equation}
\small
    C_f \equiv \frac{\tau_w}{\frac{1}{2}\rho U_b^2},~~~\tau_w \equiv \left.\mu \left( \frac{{\rm d} \langle U \rangle}{{\rm d}n} \right)\right|_w,
    \label{eq:cf}
\end{equation}
as a function of the $x$ coordinate, where $\tau_w$ is the wall-shear stress, $n$ is the wall-normal direction, and $w$ denotes the wall. 
Again, we do not see significant differences among the three grids. 
The two RANS models over-predict the size of the recirculation zone and under-predict the peak skin friction value at the windward side of the hill on all three grids.
Figure \ref{fig:pherr} quantifies the error in the skin friction as a function of the grid resolution:
\begin{equation}
\small
    {\rm err}=\sqrt{\frac{\int (C_{\rm f,pred}-C_{\rm f,DNS})^2{\rm d}x}{\int C^2_{\rm f,DNS}{\rm d}x}}\times 100,
    \label{eq:pherr}
\end{equation}
where $C_{\rm f, pred}$ is the predicted skin friction coefficient, $C_{\rm f, DNS}$ is the DNS predicted skin friction coefficient. 
We see from Fig.\ref{fig:pherr} that the RANS solution barely improves from the R-grid to the LR- and the L-grids.
The LES solution, i.e., WALE in Fig.\ref{fig:pherr}, outperforms the RANS solution on all 3 grids.
In fact, the LES solution is twice as accurate as of the RANS solution on the R-grid: the error in the R-grid LES is about 17\%, and the error in the two R-grid RANSs is about 35\%.
It is worth noting that the R-grid LES result is not grid-converged.
This should be no surprise: we cannot expect grid-converged LES on a RANS grid.
{Also important is that the RANS solution is insensitive to the grid refinement, and extra grid resolution offered no improvement in the solution.}

\subsection{Backward-Facing Step}\label{sec:bfs}

\begin{figure}
    \centering
    \includegraphics[width=0.48\textwidth]{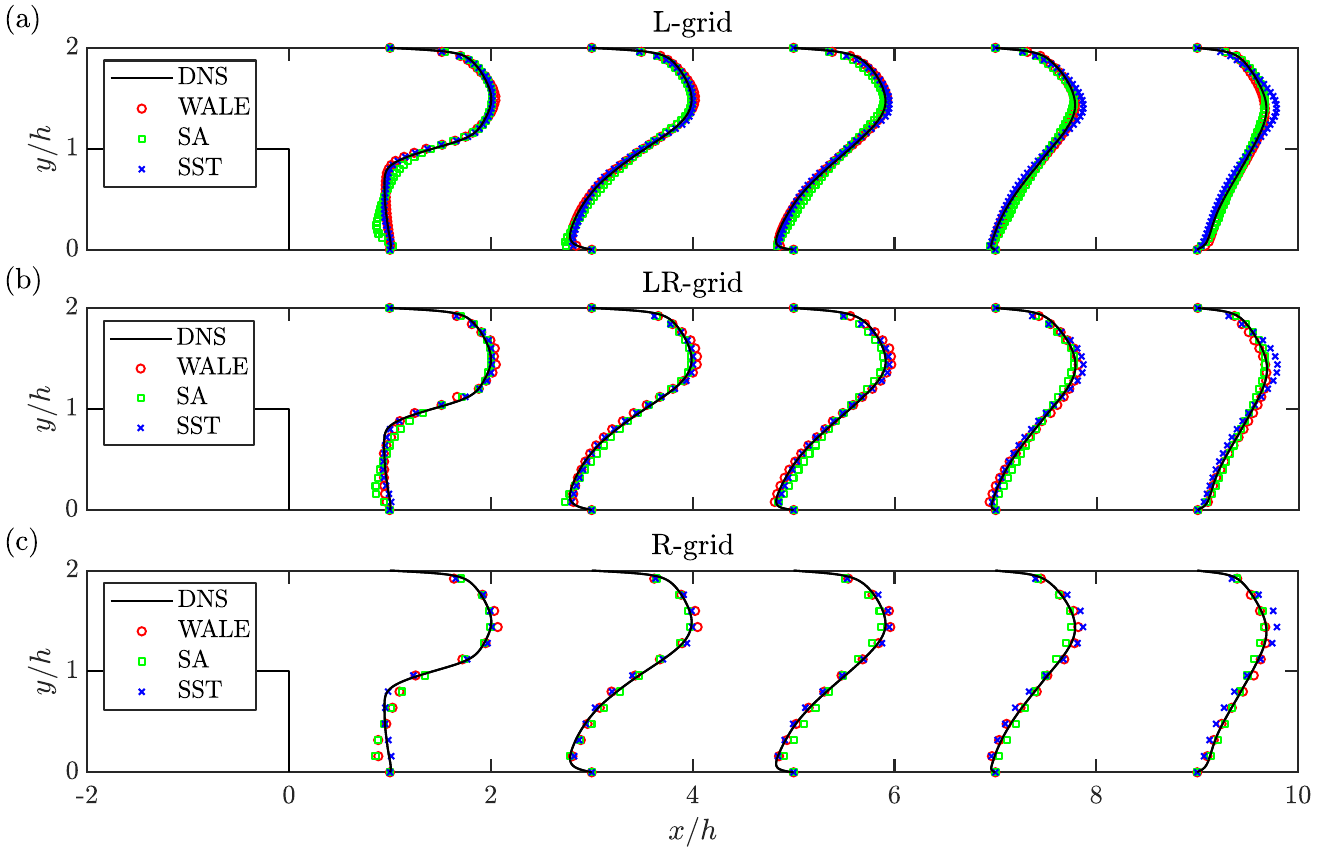}
    \caption{\label{fig:u-bfs} Same as Fig.\ref{fig:u-ph}, but for back facing step.}
\end{figure}


\begin{figure}
  \centering
  \includegraphics[width=0.24\textwidth]{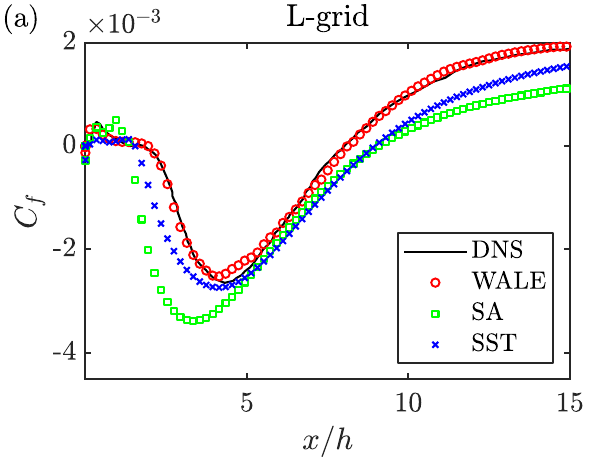}\includegraphics[width=0.24\textwidth]{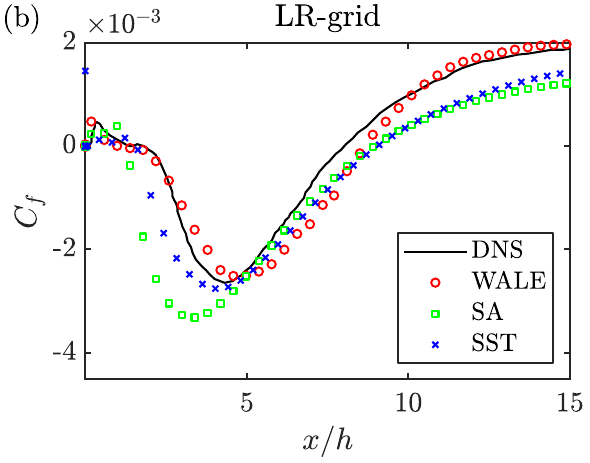}
  \includegraphics[width=0.24\textwidth]{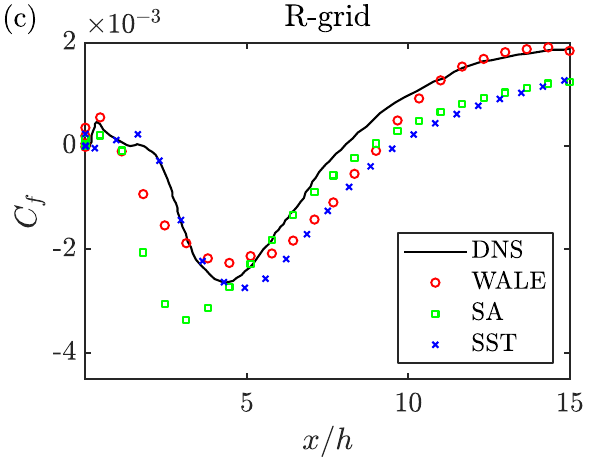}
  \caption{Skin fiction coefficient as a function of the $x$ coordinate for the back facing step case. (a) L-grid results. (b) LR-grid results. (c) R-grid results.}
  \label{fig:bfsCf}
\end{figure}

\begin{figure}
    \centering
    \includegraphics[width=0.27\textwidth]{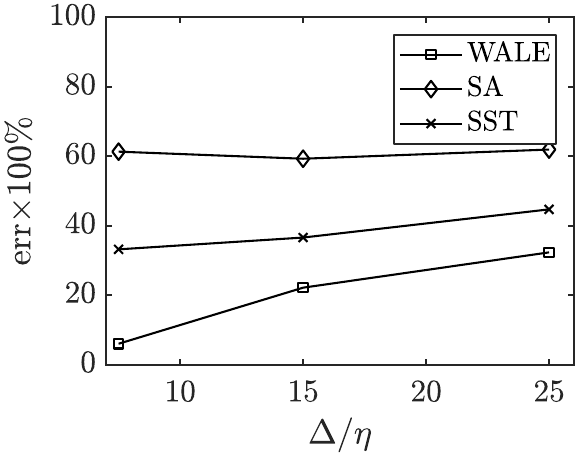}
    \caption{Errors in the LES and RANS predicted skin friction coefficient as measured by Eq. \ref{eq:pherr}.}
\label{fig:bfserr}
\end{figure}

Figure \ref{fig:u-bfs} shows the velocity profiles at a few $x$ locations in the RANSs and the LESs on the three grids.
SA is noticeably off, but both SST and WALE give reasonably accurate results on the three grids.
Figure \ref{fig:bfsCf} shows the skin friction coefficient.
Both RANS models predict a smaller skin friction coefficient, although the $k$-$\omega$ SST model seems to be slightly more accurate than the SA model.
The LES solution improves from the R-grid to the LR-grid and the L-grid, and it is more accurate than the RANS solution on all 3 grids.
Figure \ref{fig:bfserr} shows the error in the skin friction as a function of the grid resolution.
We observe the following.
Firstly, refining the R-grid does not offer much improvement for RANS.
Secondly, the LES solution, i.e., WALE in Fig.\ref{fig:bfserr}, is more accurate than the RANS solutions on all 3 grids.
Specifically, on the R-grid, the errors in the LES solution and the SA, SST RANS solutions are 32\%, 62\% and 45\%, respectively.
Thirdly, the LES solution is not grid converged on the R-grid (RANS grid).
Specifically, the errors of LES reduce from 32\% on the R grid to 22\% and 6\% on the L-grid and LR-grid, respectively.

\subsection{Jet in cross flow}\label{sec:sj}

\begin{figure}
    \centering
    \includegraphics[width=0.48\textwidth]{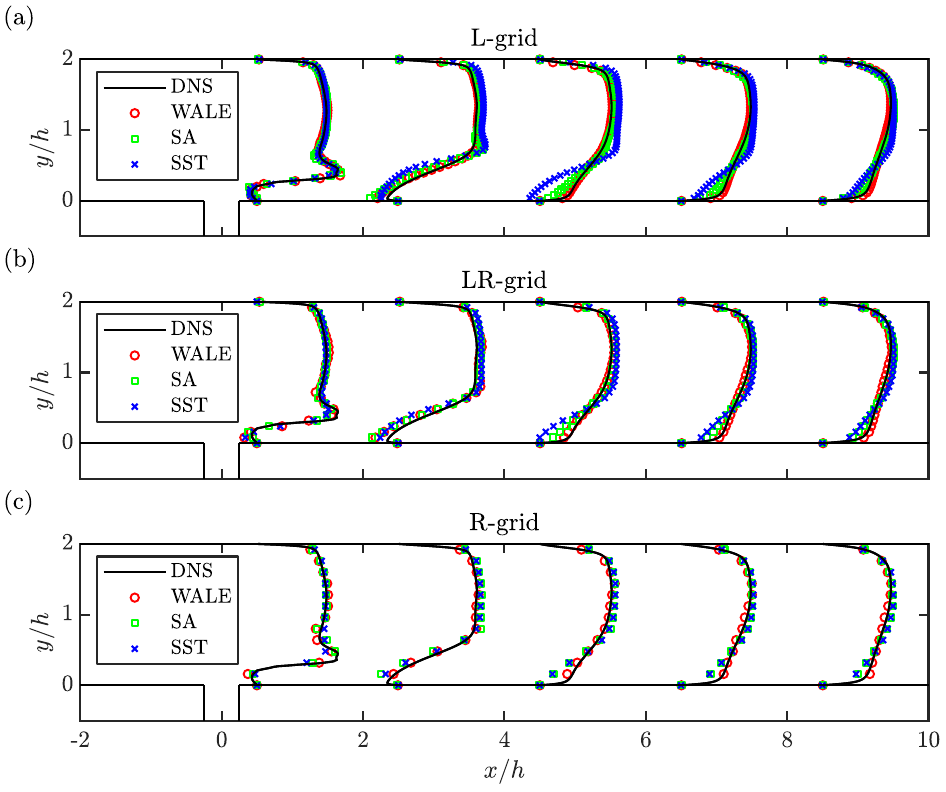}
    \caption{\label{fig:u-sj} Same as Fig.\ref{fig:u-ph}, but for jet in cross flow.}
\end{figure}


\begin{figure}[htb]
  \centering
  \includegraphics[width=0.24\textwidth]{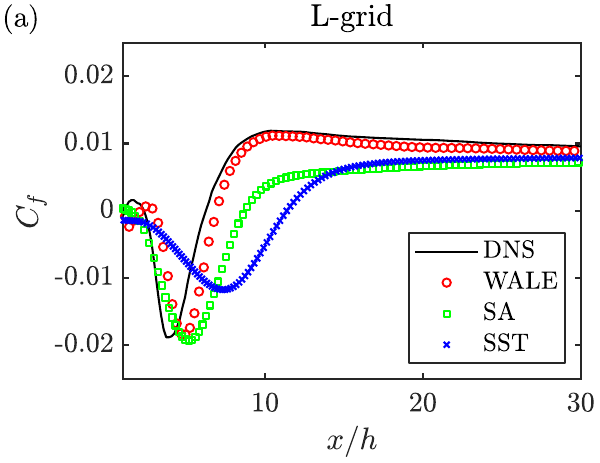}\includegraphics[width=0.24\textwidth]{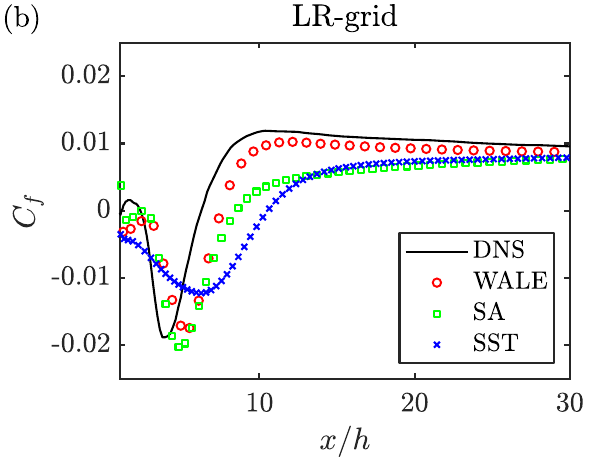}
  \includegraphics[width=0.24\textwidth]{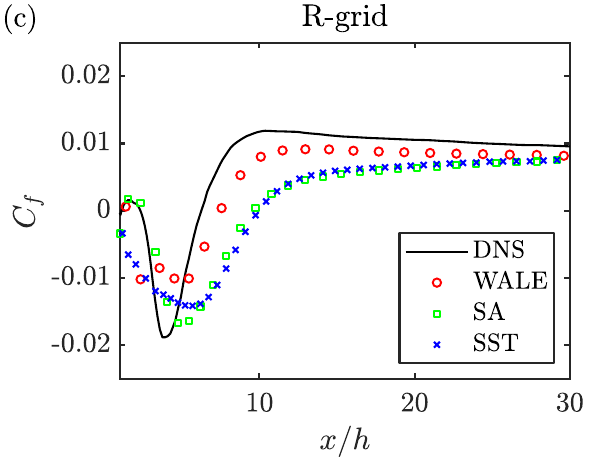}
  \caption{Same as Fig.\ref{fig:bfsCf} but for jet in cross flow.
  }
  \label{fig:sjCf}
\end{figure}

\begin{figure}[htb]
  \centering
  \includegraphics[width=0.27\textwidth]{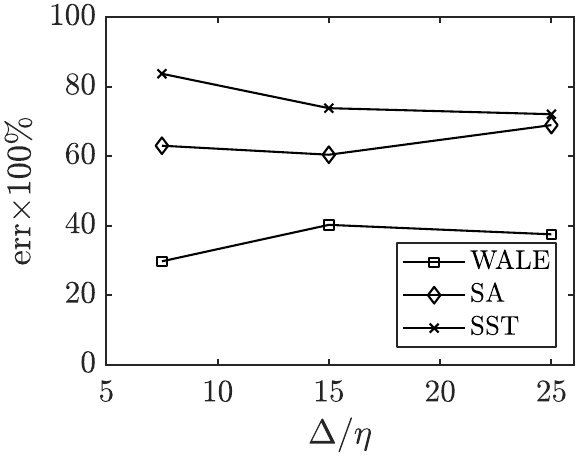}
  \caption{Same as Fig.\ref{fig:bfserr} but for jet in cross flow.}
  \label{fig:sjerr}
\end{figure}

{Figure \ref{fig:u-sj}} shows the streamwise velocity profiles at a few $x$ locations downstream of the leakage jet on the three grids.
The two RANS models predict a much more extended separation bubble, whereas the WALE model agrees fairly well with the DNS on all three grids. 
Figure \ref{fig:sjCf} shows the skin friction coefficient as a function of the $x$ coordinate for all 3 grids.
The RANSs have difficulty handling separation, which is expected.
The SA model is notably more accurate than the $k$-$\omega$ SST model, unlike the back-facing step case, for which the $k$-$\omega$ SST model is more accurate.
The LES solution follows the DNS more closely than the two RANSs on all 3 grids, and the results improve as the grid refines.
{On the contrary, the RANS results seem to have grid converged in all grids, and grid refinement offers little improvement in the RANS solution. }
Figure \ref{fig:sjerr} shows the error as a function of the grid resolution.
The error in the SA RANS slightly decreases from the R-grid to the LR-grid and stays constant between the LR-grid and the L-grid.
The error in the SST model increases as we refine the grid.
We can attribute this behavior to grid-induced separation.
On the R-grid, the LES incurs half as much error as the two RANS: the error in the LES is about 38\%, and the error in the two RANSs is about 70\%.
Again, the LES solution is not grid converged on the R-grid, and the solution keeps improving as we refine the grid.
Nonetheless, this case is challenging, and even the L-grid LES incurs a 30\% error.

\section{Discussion}
\label{sect:discussion}

\begin{figure}
  \centering
  \includegraphics[width=0.48\textwidth]{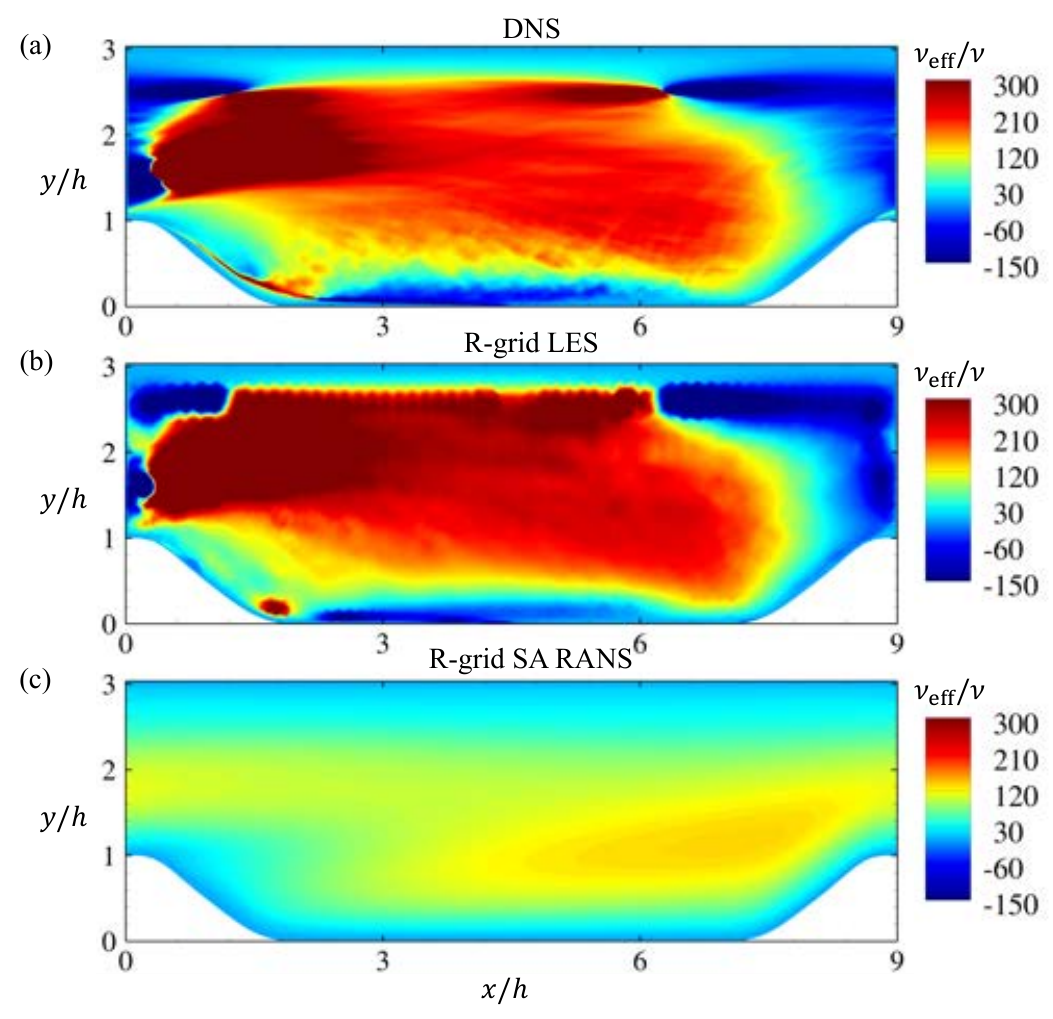}
  \caption{Time-averaged effective eddy viscosity in the periodic hill case on the R-grid.}
  \label{fig:nut}
\end{figure}

We explain why LESs on RANS-quality grids outperform RANSs on the same grids.
When the grid is fixed, the only difference between LES and RANS is how the turbulent stresses are modeled.
Figure \ref{fig:nut} shows effective eddy viscosity in DNS, LES, and RANS of the periodic hill on the R-grid.
Here, the effective eddy viscosity is 
\begin{equation}
\small
    \nu_{\rm eff}=\frac{b_{ij}\<S_{ij}\>}{2\<S_{ij}\>\<S_{ij}\>},
\end{equation}
where $b_{ij}$ is the anisotropic part of the Reynolds stress, and it is fully resolved in DNS, partly resolved in LES, and entirely modeled in RANS.  
Per the definition above, $b_{ij}$ in LES is the sum of the resolved Reynolds stress and the modeled sub-grid scale Reynolds stress, $\nu_{\rm eff}$ in RANS is the eddy viscosity $\nu_t$, and $\nu_{\rm eff}$ in DNS and LES is the effective eddy viscosity that minimizes the difference between the $2\nu_{\rm eff} \left<S_{ij}\right>$ and $b_{ij}$.
We see that the LES resembles DNS, whereas RANS significantly under-predicts the effective eddy viscosity.
Being able to capture the effective eddy viscosity is why LES outperforms RANS on RANS-quality grids. 

\begin{figure}[htb!]
  \centering
  \includegraphics[width=0.4\textwidth]{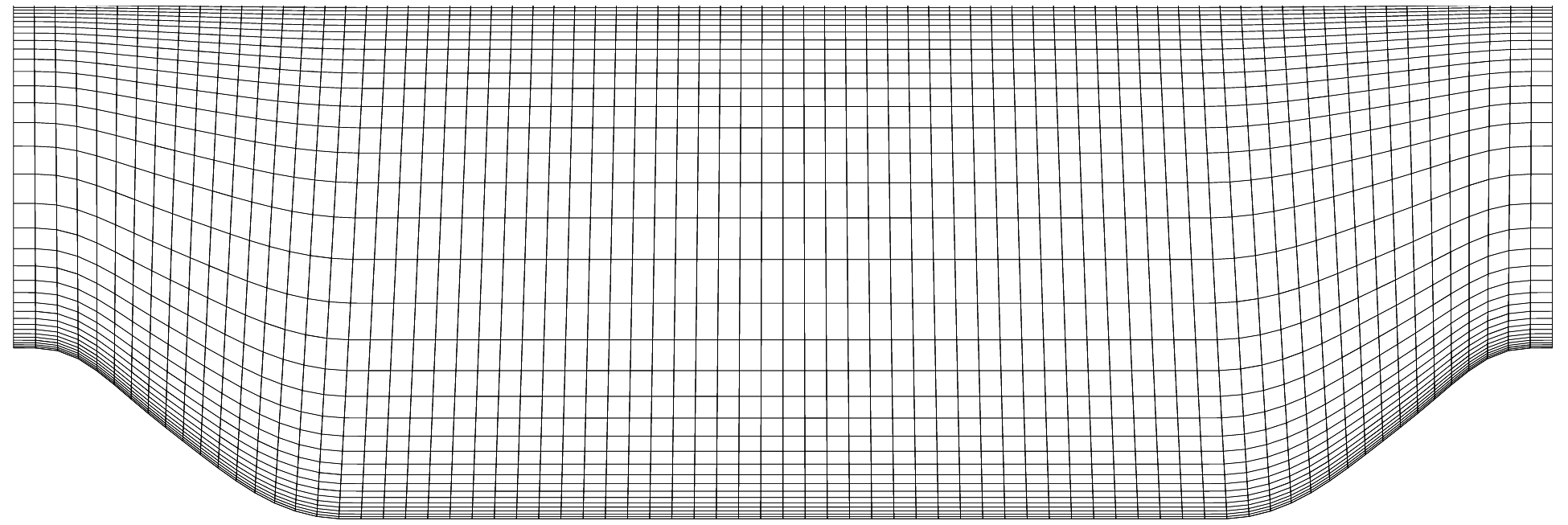}
  \caption{The RANS grid used in Ref. \cite{xiao2012consistent} for the periodic hill case.
  The grid is a structured one and its size is $74\times 37$.}
  \label{fig:trad-grid}
\end{figure}

\begin{figure}[htb!]
  \centering
  \includegraphics[width=0.27\textwidth]{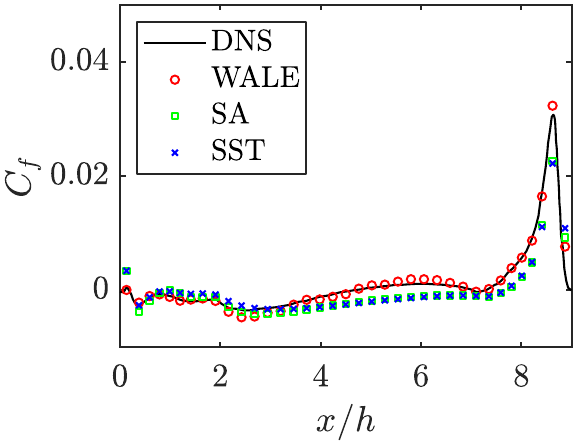}
  \caption{Skin friction prediction on the grid in Fig.\ref{fig:trad-grid}.
  The DNS results are reported in \cite{xiao2012consistent}.}
  \label{fig:trad-cf}
\end{figure}

Next, we show that the Voronoi grid is not essential and also that LES on the RANS grid outperforms RANS at higher Reynolds numbers as well.
To show these, we conduct LES and RANS of the periodic hill case at the Reynolds number $Re_h=10595$.
We use the structured grid in {Ref.} \cite{xiao2012consistent}.
The size of the grid is $74\times 37=2738$, which is slightly larger than our R-grid{, as shown in Figure \ref{fig:trad-grid}}.
Figure \ref{fig:trad-cf} shows the skin friction coefficient.
The result looks essentially the same as the one in Fig. \ref{fig:phCf} (c), and WALE outperforms the two RANS models.
Further benchmarking at high Reynolds numbers is not pursued here and is left for future work.

\section{Concluding remarks}
\label{sect:conclusions}

We explore the concept of LES on the RANS-quality grid.
LES on the RANS-quality grid differs from DES, VLES, and hybrid RANS/LES in that it does not try to match the grid and the equations.
Rather, to conduct an LES on a RANS-quality grid is to solve the LES equations on a RANS grid without switching between the RANS and LES equations.
The concept has seen successes \cite{xu2021comparative, park2016wall, goc2021largeJ} but has not been formally benchmarked for canonical cases.
This study benchmarks LES on a RANS-quality grid for three cases, namely, periodic hill, backing facing step, and jet in cross flow.
The three flows feature flow separation. 
Special attention is given to grid generation.
We use prism layers and Voronoi cells near and away from the wall, respectively.
LES and RANS grids are generated by systematically controlling the local grid spacing.
We show that for a RANS grid that no longer offers much improvement for RANS with further grid refinement, solving LES equations on that grid is highly beneficial.
Specifically, LES on the RANS-quality grid is twice as accurate as RANS in terms of predicting the skin friction coefficient.
{An implication of this is that when conducting hybrid RANS LES, where RANS is used for attached flow regions and LES is used for separated flow regions, one can get reasonable results by using coarse grids in the LES part of the simulation, particularly when it is too costly to employ fine grids in the LES part. }
The success of LES on the RANS-quality grid is attributed to its ability to capture effective eddy viscosity.
The relative cost of an LES and a RANS on the same grid roughly scales as the time steps required for the two.
For the three flows studied in this work, it takes about 7 flow-throughs to arrive at a converged RANS solution and about 30 flow-throughs to arrive at a statistically stationary state and get converged LES statistics. 
It follows that LES on a RANS-quality grid is about 5 times as expensive as RANS on the same grid, which is considerably more cost-effective than a proper LES.
 

\section*{Acknowledgments}
Bin acknowledges NNSFC grant number 91752202.
Yang acknowledges NSF for financial support.


\nocite{*}

\bibliographystyle{asmeconf}  
\bibliography{a-revision}

\begin{thebibliography}{10}
\newcommand{\enquote}[1]{``#1''}
\providecommand{\url}[1]{\texttt{#1}}
\providecommand{\urlprefix}{URL }
\expandafter\ifx\csname urlstyle\endcsname\relax
  \providecommand{\doi}[1]{DOI \discretionary{}{}{}#1}\else
  \providecommand{\doi}{DOI \discretionary{}{}{}\begingroup \urlstyle{rm}\Url}\fi
\providecommand{\eprint}[2][]{\urlprefix\url{#1#2}}

\bibitem{pope2000turbulent}
Pope, Stephen~B.
\newblock \textit{Turbulent flows}.
\newblock Cambridge university press (2000).

\bibitem{goc2021largeJ}
Goc, Konrad~A, Lehmkuhl, Oriol, Park, George~Ilhwan, Bose, Sanjeeb~T and Moin, Parviz.
\newblock \enquote{Large eddy simulation of aircraft at affordable cost: a milestone in computational fluid dynamics.}
\newblock \textit{Flow} Vol.~1 (2021).

\bibitem{zamiri2020large}
Zamiri, Ali, You, Sung~Jin and Chung, Jin~Taek.
\newblock \enquote{Large eddy simulation of unsteady turbulent flow structures and film-cooling effectiveness in a laidback fan-shaped hole.}
\newblock \textit{Aerosp Sci Technol} Vol. 100 (2020): p. 105793.

\bibitem{cheng2021large}
Cheng, Yuzhou, Jin, Tai, Luo, Kun, Li, Zongyan, Wang, Haiou and Fan, Jianren.
\newblock \enquote{Large eddy simulations of spray combustion instability in an aero-engine combustor at elevated temperature and pressure.}
\newblock \textit{Aerosp Sci Technol} Vol. 108 (2021): p. 106329.

\bibitem{zhao2019large}
Zhao, Majie, Bian, Yifan, Li, Qinling and Ye, Taohong.
\newblock \enquote{Large eddy simulation of transverse single/double jet in supersonic crossflow.}
\newblock \textit{Aerosp Sci Technol} Vol.~89 (2019): pp. 31--45.

\bibitem{slotnick2014cfd}
Slotnick, Jeffrey~P, Khodadoust, Abdollah, Alonso, Juan, Darmofal, David, Gropp, William, Lurie, Elizabeth and Mavriplis, Dimitri~J.
\newblock \enquote{{CFD} vision 2030 study: a path to revolutionary computational aerosciences.} (2014).

\bibitem{choi2012grid}
Choi, Haecheon and Moin, Parviz.
\newblock \enquote{Grid-point requirements for large eddy simulation: Chapman’s estimates revisited.}
\newblock \textit{Phys. Fluids} Vol.~24 No.~1 (2012): p. 011702.

\bibitem{yang2021grid}
Yang, Xiang~IA and Griffin, Kevin~P.
\newblock \enquote{Grid-point and time-step requirements for direct numerical simulation and large-eddy simulation.}
\newblock \textit{Phys. Fluids} Vol.~33 No.~1 (2021): p. 015108.

\bibitem{park2016wall}
Park, GI and Moin, P.
\newblock \enquote{Wall-modeled {LES}: {R}ecent applications to complex flows.}
\newblock \textit{Annual Research Briefs}  (2016): pp. 39--50.

\bibitem{goc2021large}
Goc, Konrad, Bose, Sanjeeb and Moin, Parviz.
\newblock \enquote{Large {E}ddy {S}imulation of a {R}ealistic {A}ircraft {C}onfiguration.}
\newblock \textit{Bulletin of the American Physical Society} Vol.~66 (2021).

\bibitem{xu2021comparative}
Xu, Haosen, Yang, Xiang~I, Jariwala, Vishal and Larosiliere, Louis.
\newblock \enquote{Comparative aerodynamic assessment of a multistage centrifugal compressor return channel based on {RANS} and {LES}.}
\newblock \textit{AIAA Scitech}: p. 0263. 2021.

\bibitem{rumsey2020reynolds}
Rumsey, Christopher~L, Lee, Henry~C and Pulliam, Thomas~H.
\newblock \enquote{Reynolds-{A}veraged {N}avier-{S}tokes {C}omputations of the {NASA} {J}uncture {F}low {M}odel Using {FUN3D} and {OVERFLOW}.}
\newblock \textit{AIAA Scitech}: p. 1304. 2020.

\bibitem{witherden2017future}
Witherden, Freddie~D and Jameson, Antony.
\newblock \enquote{Future directions in computational fluid dynamics.}
\newblock \textit{23rd AIAA Computational Fluid Dynamics Conference}: p. 3791. 2017.

\bibitem{zhou2021reynolds}
Zhou, Zhideng, Wu, Ting and Yang, Xiaolei.
\newblock \enquote{Reynolds number effect on statistics of turbulent flows over periodic hills.}
\newblock \textit{Phys. Fluids} Vol.~33 No.~10 (2021): p. 105124.

\bibitem{xiao2020flows}
Xiao, Heng, Wu, Jin-Long, Laizet, Sylvain and Duan, Lian.
\newblock \enquote{Flows over periodic hills of parameterized geometries: A dataset for data-driven turbulence modeling from direct simulations.}
\newblock \textit{Comput Fluids} Vol. 200 (2020): p. 104431.

\bibitem{chaouat2013hybrid}
Chaouat, Bruno and Schiestel, Roland.
\newblock \enquote{Hybrid RANS/LES simulations of the turbulent flow over periodic hills at high Reynolds number using the PITM method.}
\newblock \textit{Comput Fluids} Vol.~84 (2013): pp. 279--300.

\bibitem{hanjalic1998contribution}
Hanjali{\'c}, K and Jakirli{\'c}, S.
\newblock \enquote{Contribution towards the second-moment closure modelling of separating turbulent flows.}
\newblock \textit{Comput Fluids} Vol.~27 No.~2 (1998): pp. 137--156.

\bibitem{de2018use}
de~la Llave~Plata, Marta, Couaillier, Vincent and Le~Pape, Marie-Claire.
\newblock \enquote{On the use of a high-order discontinuous Galerkin method for DNS and LES of wall-bounded turbulence.}
\newblock \textit{Comput Fluids} Vol. 176 (2018): pp. 320--337.

\bibitem{hanson2019flow}
Hanson, David~R, McClain, Stephen~T, Snyder, Jacob~C, Kunz, Robert~F and Thole, Karen~A.
\newblock \enquote{Flow in a scaled turbine coolant channel with roughness due to additive manufacturing.}
\newblock \textit{Turbo Expo: Power for Land, Sea, and Air}, Vol. 58653: p. V05BT21A004. 2019. American Society of Mechanical Engineers.

\bibitem{menter2003ten}
Menter, Florian~R, Kuntz, Martin and Langtry, Robin.
\newblock \enquote{Ten years of industrial experience with the {SST} turbulence model.}
\newblock \textit{Turbulence, heat and mass transfer} Vol.~4 No.~1 (2003): pp. 625--632.

\bibitem{fortune1995voronoi}
Fortune, Steven.
\newblock \enquote{Voronoi diagrams and {D}elaunay triangulations.}
\newblock \textit{Computing in Euclidean geometry}  (1995): pp. 225--265.

\bibitem{bres2018large}
Bres, Guillaume~A, Bose, Sanjeeb~T, Emory, Michael, Ham, Frank~E, Schmidt, Oliver~T, Rigas, Georgios and Colonius, Tim.
\newblock \enquote{Large-eddy simulations of co-annular turbulent jet using a {V}oronoi-based mesh generation framework.}
\newblock \textit{2018 AIAA/CEAS Aeroacoustics Conference}: p. 3302. 2018.

\bibitem{lozano2021performance}
Lozano-Dur{\'a}n, Adri{\'a}n, Bose, Sanjeeb~T and Moin, Parviz.
\newblock \enquote{Performance of wall-modeled {LES} with boundary-layer-conforming grids for external aerodynamics.}
\newblock \textit{AIAA J.}  (2021): pp. 1--20.

\bibitem{engwirda2015voronoi}
Engwirda, Darren.
\newblock \enquote{Voronoi-based point-placement for three-dimensional Delaunay-refinement.}
\newblock \textit{Procedia Engineering} Vol. 124 (2015): pp. 330--342.

\bibitem{engwirda2018generalised}
Engwirda, Darren.
\newblock \enquote{Generalised primal-dual grids for unstructured co-volume schemes.}
\newblock \textit{Journal of Computational Physics} Vol. 375 (2018): pp. 155--176.

\bibitem{nicoud1999subgrid}
Nicoud, Franck and Ducros, Fr{\'e}d{\'e}ric.
\newblock \enquote{Subgrid-scale stress modelling based on the square of the velocity gradient tensor.}
\newblock \textit{Flow Turbul. Combust.} Vol.~62 No.~3 (1999): pp. 183--200.

\bibitem{menter1994two}
Menter, Florian~R.
\newblock \enquote{Two-equation eddy-viscosity turbulence models for engineering applications.}
\newblock \textit{AIAA J.} Vol.~32 No.~8 (1994): pp. 1598--1605.

\bibitem{spalart1992one}
Spalart, Philippe and Allmaras, Steven.
\newblock \enquote{A one-equation turbulence model for aerodynamic flows.}
\newblock \textit{30th aerospace sciences meeting and exhibit}: p. 439. 1992.

\bibitem{zhou2021wall}
Zhou, Zhideng, He, Guowei and Yang, Xiaolei.
\newblock \enquote{Wall model based on neural networks for {LES} of turbulent flows over periodic hills.}
\newblock \textit{Phys. Rev. Fluids} Vol.~6 No.~5 (2021).

\bibitem{xie2008efficient}
Xie, Zheng-Tong and Castro, Ian~P.
\newblock \enquote{Efficient generation of inflow conditions for large eddy simulation of street-scale flows.}
\newblock \textit{Flow Turbul. Combust.} Vol.~81 No.~3 (2008): pp. 449--470.

\bibitem{krank2018direct}
Krank, Benjamin, Kronbichler, Martin and Wall, Wolfgang~A.
\newblock \enquote{Direct {N}umerical {S}imulation of {F}low over {P}eriodic {H}ills up to $Re_H=10,595$.}
\newblock \textit{Flow Turbul. Combust.} Vol. 101 No.~2 (2018): pp. 521--551.

\bibitem{barri2009dns}
Barri, Mustafa, El~Khoury, George~K., Andersson, Helge~I. and Pettersen, Bjørnar.
\newblock \enquote{{DNS} of backward-facing step flow with fully turbulent inflow.}
\newblock \textit{Int J Numer Methods Fluids}  (2009): p. 105106.

\bibitem{xu2022direct}
Xu, H H~A, Lynch, Stephen and Yang, X I~A.
\newblock \enquote{Direct numerical simulation of slot film cooling downstream of misaligned plates.}
\newblock \textit{Flow} Vol.~2 (2022): p.~E7.

\bibitem{balakumar2014dns}
Balakumar, P, Park, GI and Pierce, B.
\newblock \enquote{{DNS}, {LES}, and wall-modeled {LES} of separating flow over periodic hills.}
\newblock \textit{Proceedings of the Summer Program}: pp. 407--415. 2014.

\bibitem{xiao2012consistent}
Xiao, Heng and Jenny, Patrick.
\newblock \enquote{A consistent dual-mesh framework for hybrid {LES/RANS} modeling.}
\newblock \textit{J Comput Phys} Vol. 231 No.~4 (2012): pp. 1848--1865.

\bibitem{juste2016assessment}
Juste, GL, Fajardo, P and Guijarro, A.
\newblock \enquote{Assessment of secondary bubble formation on a backward-facing step geometry.}
\newblock \textit{Phys. Fluids} Vol.~28 No.~7 (2016): p. 074106.

\bibitem{wu2013mixed}
Wu, Xin, Ju, Ping and Wu, Feng.
\newblock \enquote{A mixed-time-scale low-{R}eynolds-number one-equation turbulence model.}
\newblock \textit{J. Turbul} Vol.~14 No.~4 (2013): pp. 55--87.

\bibitem{park2017wall}
Park, George~Ilhwan.
\newblock \enquote{Wall-modeled large-eddy simulation of a high Reynolds number separating and reattaching flow.}
\newblock \textit{AIAA Journal} Vol.~55 No.~11 (2017): pp. 3709--3721.

\bibitem{park2016space}
Park, George~Ilhwan and Moin, Parviz.
\newblock \enquote{Space-time characteristics of wall-pressure and wall shear-stress fluctuations in wall-modeled large eddy simulation.}
\newblock \textit{Physical review fluids} Vol.~1 No.~2 (2016): p. 024404.

\bibitem{piomelli2002wall}
Piomelli, Ugo and Balaras, Elias.
\newblock \enquote{Wall-layer models for large-eddy simulations.}
\newblock \textit{Annual review of fluid mechanics} Vol.~34 No.~1 (2002): pp. 349--374.

\bibitem{laurence2002large}
Laurence, Dominique.
\newblock \enquote{Large {E}ddy {S}imulation of {I}ndustrial {F}lows?}
\newblock \textit{Closure Strategies for Turbulent and Transitional Flows, ed. {\rm BE Launder, ND Sandham}}.
\newblock Cambridge Univ. Press (2002): pp. 392--406.

\bibitem{schaller1997moore}
Schaller, Robert~R.
\newblock \enquote{Moore's law: past, present and future.}
\newblock \textit{IEEE spectrum} Vol.~34 No.~6 (1997): pp. 52--59.

\bibitem{mack2011fifty}
Mack, Chris~A.
\newblock \enquote{Fifty years of {M}oore's law.}
\newblock \textit{IEEE Trans Compon Packaging Manuf Technol} Vol.~24 No.~2 (2011): pp. 202--207.

\bibitem{zhao2020bypass}
Zhao, Yaomin and Sandberg, Richard~D.
\newblock \enquote{Bypass transition in boundary layers subject to strong pressure gradient and curvature effects.}
\newblock \textit{J. Fluid Mech.} Vol. 888 (2020).

\bibitem{zhao2021high}
Zhao, Yaomin and Sandberg, Richard~D.
\newblock \enquote{High-fidelity simulations of a high-pressure turbine vane subject to large disturbances: effect of exit Mach number on losses.}
\newblock \textit{J Turbomach} Vol. 143 No.~9 (2021).

\bibitem{hanjalic2005will}
Hanjalic, K.
\newblock \enquote{{Will {RANS} {S}urvive {LES}? {A} {V}iew of {P}erspectives}.}
\newblock \textit{J. Fluids Eng.} Vol. 127 No.~5 (2005): pp. 831--839.

\bibitem{wilcox1998turbulence}
Wilcox, David~C et~al.
\newblock \textit{Turbulence modeling for {CFD}}.
\newblock Vol.~2.
\newblock DCW industries La Canada, CA (1998).

\bibitem{bogard2006gas}
Bogard, David~G and Thole, Karen~A.
\newblock \enquote{Gas turbine film cooling.}
\newblock \textit{J. Propuls. Power} Vol.~22 No.~2 (2006): pp. 249--270.

\bibitem{stevens2014concurrent}
Stevens, Richard~JAM, Graham, Jason and Meneveau, Charles.
\newblock \enquote{A concurrent precursor inflow method for large eddy simulations and applications to finite length wind farms.}
\newblock \textit{Renew. Energ.} Vol.~68 (2014): pp. 46--50.

\bibitem{chen2012reynolds}
Chen, Shiyi, Xia, Zhenhua, Pei, Suyang, Wang, Jianchun, Yang, Yantao, Xiao, Zuoli and Shi, Yipeng.
\newblock \enquote{Reynolds-stress-constrained large-eddy simulation of wall-bounded turbulent flows.}
\newblock \textit{J. Fluid Mech.} Vol. 703 (2012): pp. 1--28.

\bibitem{kim1987turbulence}
Kim, John, Moin, Parviz and Moser, Robert.
\newblock \enquote{Turbulence statistics in fully developed channel flow at low {R}eynolds number.}
\newblock \textit{J. Fluid Mech.} Vol. 177 (1987): pp. 133--166.

\bibitem{weller1998tensorial}
Weller, Henry~G, Tabor, Gavin, Jasak, Hrvoje and Fureby, Christer.
\newblock \enquote{A tensorial approach to computational continuum mechanics using object-oriented techniques.}
\newblock \textit{Comput Phys} Vol.~12 No.~6 (1998): pp. 620--631.

\bibitem{xu2021flow}
Xu, Haosen~HA, Altland, Samuel~J, Yang, Xiang~IA and Kunz, Robert~F.
\newblock \enquote{Flow over closely packed cubical roughness.}
\newblock \textit{J. Fluid Mech.} Vol. 920 (2021).

\bibitem{tracy2020large}
Tracy, Kevin and Lynch, Stephen~P.
\newblock \enquote{Large eddy simulation of the 7-7-7 shaped film cooling hole at axial and compound angle orientations.}
\newblock \textit{Turbo Expo: Power for Land, Sea, and Air}, Vol. 84171: p. V07BT12A016. 2020. American Society of Mechanical Engineers.

\bibitem{lozano2020non}
Lozano-Dur{\'a}n, Adri{\'a}n, Giometto, Marco~G, Park, George~Ilhwan and Moin, Parviz.
\newblock \enquote{Non-equilibrium three-dimensional boundary layers at moderate Reynolds numbers.}
\newblock \textit{J. Fluid Mech.} Vol. 883 (2020).

\bibitem{abkar2016wake}
Abkar, Mahdi, Sharifi, Ahmad and Port{\'e}-Agel, Fernando.
\newblock \enquote{Wake flow in a wind farm during a diurnal cycle.}
\newblock \textit{J. Turbul} Vol.~17 No.~4 (2016): pp. 420--441.

\bibitem{giometto2017effects}
Giometto, Marco~Giovanni, Christen, Andreas, Egli, Pascal~Emanuel, Schmid, MF, Tooke, RT, Coops, NC and Parlange, Marc~B.
\newblock \enquote{Effects of trees on mean wind, turbulence and momentum exchange within and above a real urban environment.}
\newblock \textit{Adv Water Resour} Vol. 106 (2017): pp. 154--168.

\bibitem{guo2022practical}
Guo, Xianwen, Xia, Zhenhua and Chen, Shiyi.
\newblock \enquote{Practical framework for data-driven {RANS} modeling with data augmentation.}
\newblock \textit{Acta Mech Sin}  (2022): pp. 1--9.

\bibitem{thangam1992turbulent}
Thangam, S and Speziale, Charles~G.
\newblock \enquote{Turbulent flow past a backward-facing step-A critical evaluation of two-equation models.}
\newblock \textit{AIAA J.} Vol.~30 No.~5 (1992): pp. 1314--1320.

\bibitem{xu2021slot}
Xu, Haosen, Lynch, Stephen and Yang, Xiang.
\newblock \enquote{Slot film cooling downstream of misaligned plates.}
\newblock \textit{Bulletin of the American Physical Society} Vol.~66 (2021).

\bibitem{xu2021assessing}
Xu, Haosen~HA, Yang, Xiang~IA and Milani, Pedro~M.
\newblock \enquote{Assessing wall-modeled large-eddy simulation for low-speed flows with heat transfer.}
\newblock \textit{AIAA J.} Vol.~59 No.~6 (2021): pp. 2060--2069.

\bibitem{spalart2009detached}
Spalart, Philippe~R.
\newblock \enquote{Detached-eddy simulation.}
\newblock \textit{Ann. Rev. Fluid Mech.} Vol.~41 (2009): pp. 181--202.

\bibitem{speziale1998turbulence}
Speziale, Charles~G.
\newblock \enquote{Turbulence modeling for time-dependent {RANS} and {VLES}: a review.}
\newblock \textit{AIAA J.} Vol.~36 No.~2 (1998): pp. 173--184.

\bibitem{durbin2018some}
Durbin, Paul~A.
\newblock \enquote{Some recent developments in turbulence closure modeling.}
\newblock \textit{Ann. Rev. Fluid Mech.} Vol.~50 (2018): pp. 77--103.

\bibitem{liu2006turbulence}
Liu, Nan-Suey and Shih, Tsan-Hsing.
\newblock \enquote{Turbulence modeling for very large-eddy simulation.}
\newblock \textit{AIAA J.} Vol.~44 No.~4 (2006): pp. 687--697.

\bibitem{bose2018wall}
Bose, Sanjeeb~T and Park, George~Ilhwan.
\newblock \enquote{Wall-modeled large-eddy simulation for complex turbulent flows.}
\newblock \textit{Ann. Rev. Fluid Mech.} Vol.~50 (2018): pp. 535--561.

\bibitem{christopher2019fun3d}
Rumsey, Christopher~L., Carlson, Jan and Ahmad, Nashat.
\newblock \enquote{{FUN3D} {J}uncture {F}low {C}omputations {C}ompared with {E}xperimental {D}ata.} (2019).

\bibitem{michael2019an}
Kegerise, Michael~A, Neuhart, Dan, Hannon, Judith and Rumsey, Christopher~L.
\newblock \enquote{An experimental investigation of a wing-fuselage junction model in the {NASA} Langley 14-by 22-foot Subsonic Wind Tunnel.}
\newblock \textit{AIAA Scitech 2019 Forum}. Conference proceedings: p. 0077. 2019.

\bibitem{batten2004interfacing}
Batten, Paul, Goldberg, Uriel and Chakravarthy, Sukumar.
\newblock \enquote{Interfacing statistical turbulence closures with large-eddy simulation.}
\newblock \textit{AIAA J.} Vol.~42 No.~3 (2004): pp. 485--492.

\bibitem{ruprecht2002simulation}
Ruprecht, Albert, Helmrich, Thomas, Aschenbrenner, Thomas and Scherer, Thomas.
\newblock \enquote{Simulation of vortex rope in a turbine draft tube.}
\newblock \textit{Proceedings of the 21st IAHR Symposium on Hydraulic Machinery and Systems}, Vol.~1: pp. 259--266. 2002. EPFL/STI/LMH, Lausanne, Switzerland.

\bibitem{chen2018review}
Chen, Lin, Asai, Keisuke, Nonomura, Taku, Xi, Guannan and Liu, Tianshu.
\newblock \enquote{A review of {B}ackward-{F}acing {S}tep ({BFS}) flow mechanisms, heat transfer and control.}
\newblock \textit{Therm. Sci. Eng. Prog.} Vol.~6 (2018): pp. 194--216.

\bibitem{rizzetta2002application}
Rizzetta, Donald~P and Visbal, Miguel~R.
\newblock \enquote{Application of large-eddy simulation to supersonic compression ramps.}
\newblock \textit{AIAA J.} Vol.~40 No.~8 (2002): pp. 1574--1581.

\bibitem{raverdy2003high}
Raverdy, B, Mary, I, Sagaut, P and Liamis, N.
\newblock \enquote{High-resolution large-eddy simulation of flow around low-pressure turbine blade.}
\newblock \textit{AIAA J.} Vol.~41 No.~3 (2003): pp. 390--397.

\bibitem{michelassi2002analysis}
Michelassi, Vittorio, Wissink, Jan and Rodi, Wolfgang.
\newblock \enquote{Analysis of {DNS} and {LES} of flow in a low pressure turbine cascade with incoming wakes and comparison with experiments.}
\newblock \textit{Flow Turbul. Combust.} Vol.~69 No.~3 (2002): pp. 295--329.

\bibitem{spalart2015direct}
Spalart, Phillipe~R, Shur, Michael~L, Strelets, M~Kh and Travin, Andrey~K.
\newblock \enquote{Direct simulation and {RANS} modelling of a vortex generator flow.}
\newblock \textit{Flow Turbul. Combust.} Vol.~95 No.~2 (2015): pp. 335--350.

\bibitem{shur2011noise}
Shur, Michael~L, Spalart, Philippe~R and Strelets, Michael~Kh.
\newblock \enquote{Noise prediction for underexpanded jets in static and flight conditions.}
\newblock \textit{AIAA J.} Vol.~49 No.~9 (2011): pp. 2000--2017.

\bibitem{schumann2020assessment}
Schumann, Jan-Erik, Toosi, Siavash and Larsson, Johan.
\newblock \enquote{Assessment of grid anisotropy effects on large-eddy-simulation models with different length scales.}
\newblock \textit{AIAA J.} Vol.~58 No.~10 (2020): pp. 4522--4533.

\bibitem{carton2014assessment}
Carton~de Wiart, Corentin, Hillewaert, Koen, Duponcheel, Matthieu and Winckelmans, Gr{\'e}goire.
\newblock \enquote{Assessment of a discontinuous Galerkin method for the simulation of vortical flows at high Reynolds number.}
\newblock \textit{International Journal for Numerical Methods in Fluids} Vol.~74 No.~7 (2014): pp. 469--493.

\bibitem{lv2021discontinuous}
Lv, Yu, Yang, Xiang~IA, Park, George~I and Ihme, Matthias.
\newblock \enquote{A discontinuous Galerkin method for wall-modeled large-eddy simulations.}
\newblock \textit{Computers \& Fluids} Vol. 222 (2021): p. 104933.

\bibitem{lv2018underresolved}
Lv, Yu, Ma, Peter~C and Ihme, Matthias.
\newblock \enquote{On underresolved simulations of compressible turbulence using an entropy-bounded DG method: Solution stabilization, scheme optimization, and benchmark against a finite-volume solver.}
\newblock \textit{Computers \& Fluids} Vol. 161 (2018): pp. 89--106.

\bibitem{vermeire2017utility}
Vermeire, Brian~C, Witherden, Freddie~D and Vincent, Peter~E.
\newblock \enquote{On the utility of GPU accelerated high-order methods for unsteady flow simulations: A comparison with industry-standard tools.}
\newblock \textit{J Comput Phys} Vol. 334 (2017): pp. 497--521.

\end{thebibliography}

\end{document}